\documentclass[twocolumn]{aastex62}
\usepackage{longtable}
\newcommand{\zabs}{$z_{\rm abs}$}

\newcommand{\zq}{$z_{\rm q}$}

\newcommand{\kms}{km\,s$^{-1}$}
\newcommand{\cmsq}{cm$^{-2}$}

\newcommand{\hi}{\mbox{H\,{\sc i}}}

\newcommand{\civ}{\mbox{C\,{\sc iv}}}

\def\h2{$\rm H_2$}
\def\Nh2{$N$(H${_2}$)}

\def\lya{\ensuremath{{\rm Ly}\alpha}}

\def\kms{km\,s$^{-1}$}

\def\zabs{$z_{\rm abs}$}
\def\zem{$z_{\rm em}$}

\def\21{21-cm}

\def\t0{T$_{0}$}

\def\c21{$C_{21}$}

\submitjournal{ApJ}
\shorttitle{uGMRT-SALT/NOT survey at 2$<z<$5} 
\shortauthors{Gupta et al.}
\begin{document}

\title{Evolution of cold gas at $2<z<5$: a blind search for \hi\ and OH absorption lines towards mid-infrared color selected radio-loud AGNs} 

\correspondingauthor{N. Gupta}
\email{ngupta@iucaa.in}

\author{N. Gupta},  
\affil{Inter-University Centre for Astronomy and Astrophysics, Post Bag 4, Ganeshkhind, Pune 411 007, India}

\author{R. Srianand},  
\affil{Inter-University Centre for Astronomy and Astrophysics, Post Bag 4, Ganeshkhind, Pune 411 007, India}

\author{G. Shukla},  
\affil{Inter-University Centre for Astronomy and Astrophysics, Post Bag 4, Ganeshkhind, Pune 411 007, India}

\author{J-.K. Krogager}  
\affil{Institut d'astrophysique de Paris, UMR 7095, CNRS-SU, 98bis bd Arago, 75014  Paris, France}
\affil{Department of Astronomy, University of Geneva, Chemin Pegasi 51, 1290 Versoix, Switzerland}

\author{P. Noterdaeme}  
\affil{Institut d'astrophysique de Paris, UMR 7095, CNRS-SU, 98bis bd Arago, 75014  Paris, France}

\author{F. Combes}  
\affil{ Observatoire de Paris, Coll\`ege de France, PSL University, Sorbonne University, CNRS, LERMA, Paris, France}

\author{R. Dutta}  
\affil{Dipartimento di Fisica G. Occhialini, Università degli Studi di Milano-Bicocca, Piazza della Scienza 3, 20126 Milano, Italy}

\author{J. P. U. Fynbo}  
\affil{Cosmic Dawn Center (DAWN), University of Copenhagen, Jagtvej 128, DK-2200, Copenhagen N, Denmark}
\affil{Niels Bohr Institute, University of Copenhagen, Jagtvej 128, DK-2200, Copenhagen N, Denmark}

\author{M. Hilton}  
\affil{Astrophysics Research Centre and School of Mathematics, Statistics and Computer Science, UKZN, Durban 4041, South Africa }

\author{E. Momjian}  
\affil{National Radio Astronomy Observatory, Socorro, NM 87801, USA}

\author{K. Moodley}  
\affil{Astrophysics Research Centre and School of Mathematics, Statistics and Computer Science, UKZN, Durban 4041, South Africa }

\author{P. Petitjean}  
\affil{Institut d'astrophysique de Paris, UMR 7095, CNRS-SU, 98bis bd Arago, 75014  Paris, France}

\begin{abstract}
	We present results from a spectroscopically blind search for associated and intervening \hi\ 21-cm and OH 18-cm absorption lines towards 88 AGNs at $2\le z\le5$ using the upgraded Giant Metrewave Radio Telescope (uGMRT). The sample of AGNs with 1.4\,GHz spectral luminosity in the range, $10^{27 - 29.3}$\,W\,Hz$^{-1}$,  is selected using mid-infrared colors and closely resembles the distribution of the underlying quasar population. The search for associated or proximate absorption, defined to be within 3000\,\kms\ of the AGN redshift, led to one \hi\ 21-cm absorption detection (M1540$-$1453; \zabs$=$ 2.1139). This is only the fourth known absorption at $z>2$. The detection rate ($1.6^{+3.8}_{-1.4}$\%) suggests low covering factor of cold neutral medium (CNM; T$\sim$100\,K) associated with these powerful AGNs. The intervening absorption line search, with a sensitivity to detect CNM in damped Ly$\alpha$ systems (DLAs), has comoving absorption path lengths of $\Delta$X = 130.1 and 167.7 for \hi\ and OH, respectively. The corresponding number of absorber per unit comoving path lengths are $\le$0.014 and $\le$0.011, respectively. The former is at least 4.5 times lower than that of DLAs and consistent with the CNM cross-section estimated using H$_2$ and C~{\sc i} absorbers at $z>2$. Our AGN sample is optically fainter compared to the quasars used to search for DLAs in the past. In our optical spectra obtained using the Southern African Large Telescope (SALT) and the Nordic Optical Telescope (NOT), we detect 5 intervening (redshift path$\sim9.3$) and 2 proximate DLAs.  This is slightly excessive compared to the statistics based on optically selected quasars. The non-detection of \hi\ 21-cm absorption from these DLAs suggests small CNM covering fraction around galaxies at $z>2$.
\end{abstract}  

\keywords{quasars: absorption lines ---  interstellar medium}

\section{Introduction} 
\label{sec:intro}  

\hi\ 21-cm absorption lines in the spectra of radio sources can provide valuable insights into the cold atomic gas ($T\sim$100\,K) associated with active galactic nuclei (AGNs) and intervening galaxies along the line of sight. In the former, generally detected within a few 1000\,\kms\ of the AGN redshift, the matter may be {\it associated} with the AGN, its host galaxy, a nearby companion galaxy, outflows driven by its feedback or infalling material. In the latter, the absorbing gas corresponds to the interstellar or circumgalactic medium of an {\it intervening} galaxy or intragroup medium. 
The strength of absorption signal does not depend on the distance to the observer. The \hi\ 21-cm absorption line's strength depends both on the \hi\ column density and the spin temperature of the gas. It could thus be an important probe of the properties of cold gas in distant galaxies and investigating its role in fueling the cosmic evolution of star formation rate (SFR) density and that of the luminosity density of AGNs, both of which peak at $z \simeq 2$. 

For a long time radio telescopes have receivers capable of observing the \hi\ 21-cm line  up to arbitrarily high redshifts.  Indeed, \hi\ 21-cm absorption has been searched in AGNs as distant as $z\sim 5.2$ \citep[e.g.,][]{Brown73, Carilli07, Carilli98}.  
But technical limitations imposed by narrow bandwidths, hostile radio frequency environment and limited number of known bright radio AGNs at high-$z$ have prevented large unbiased radio absorption line surveys. 
Consequently, to date, the majority of \hi\ 21-cm absorption line observations and detections have been based on optically selected samples of AGNs. 
%

For {\it associated} absorption, AGNs with known redshifts and, 
preferably with compact radio morphology, have been observed to study the circumnuclear gas which may be fueling the radio activity or impacted by the AGN feedback \citep[e.g.,][]{Vermeulen03, Gupta06, Darling11, Curran13, Allison14, Gereb15, Aditya16, Dutta19, Grasha19}.
Although more than 500 AGNs have been searched for \hi\ 21-cm absorption the vast majority of observations are at $z<2$ and most of the detections at $z<1$ \citep[see][for a review]{Morganti18}. 
Only 3 detections at $z>2$ are known, the highest redshift  being 3.53 \citep[][]{Aditya21}.
Overall, the bulk of detections are towards compact radio sources (detection rate $\sim30-50\%$) associated with galaxies having mid-infrared (MIR) colors suggesting gas and dust rich environment \citep[][]{Glowacki17, Chandola20}.  Among detections associated with more powerful AGNs (radio luminosity, log($L_{\rm 1.4\,GHz}/({\rm W\,Hz}^{-1}) > 24$), the \hi\ absorption profiles often show signatures of radio jet-ISM interaction in the form of blue-shifted components representing outflowing gas \citep[][]{Maccagni17}.
%

For {\it intervening} \hi\ 21-cm absorption line studies the targets have been sight lines towards quasars, the most powerful AGNs, selected from the large optical spectroscopic surveys such as the Sloan Digital Sky Survey \citep[SDSS;][]{York00}. 
Generally, sight lines with indications of large \hi\ column densities ($N$(\hi)) along the sight line suggested by the presence of a damped \lya\ system \citep[DLAs; $N$(\hi)$> 2\times 10^{20}$\cmsq;  e.g.,][]{Srianand12dla, Kanekar14}, a strong Mg~{\sc ii} absorption \citep[rest equivalent width, $W_{\rm r} > 1\AA$; e.g.,][]{Gupta12, Dutta17mg2}, or a galaxy at small impact parameter \citep[typially $<$30\,kpc; e.g.,][]{Carilli92, Gupta10, Borthakur10, Reeves16, Dutta17} are selected.  
The vast majority of the observations are sensitive to detecting cold neutral medium (CNM; $T\sim$100\,K) in $N(\hi) > 5 \times 10^{19}$\,\cmsq. The detection rates are typically 10-50\%, depending crucially on the sample selection criteria \citep[see, for example,][]{Dutta17fe2}.
Although the highest redshift detection is at $z\sim$3.38 \citep[][]{Kanekar07}, the bulk of the reported H~{\sc i} 21-cm detections are associated with gas rich galaxies at $z<2$. These studies also  suggest that the gas traced by DLAs at $z>2$ is predominantly warm ($T >$1000\,K).

It is reasonable to expect optically selected samples of AGNs to be affected by dust-bias.  
Since cold gas is accompanied by dust, the bias is particularly relevant for \hi\ 21-cm absorption line searches.  
In the case of associated absorption, the dust intrinsic to AGN may remove objects with certain orientation (Type\,{\sc ii}) or going through the very early stages of evolution. 
In the case of intervening gas, it can substantially affect our ability to use optically selected samples of DLAs to detect translucent and dense phases of the ISM \citep[][]{Krogager16, Geier19}, and influence the measurements of \hi\ and metal mass densities \citep[][]{Krogager19}.

The limitations due to dust obscuration can be overcome by selecting AGNs without resorting to any optical color selection scheme or carry out blind searches of \hi\ 21-cm absorption.  The latter is becoming possible with various precursor and pathfinder telescopes of Square Kilometre Array (SKA) equipped with wideband receivers.  
Especially, the upcoming large \hi\ 21-cm absorption line surveys such as the MeerKAT Absorption Line Survey \citep[MALS;][]{Gupta17mals} and First Large Absorption Survey in \hi\ \citep[FLASH;][]{Allison17} will characterize the evolution of cold gas without possible selection effects due to dust-bias or from the choice of different methods used to select sight lines in different redshift ranges \citep[see also][]{Grasha20}. 
These will also simultaneously search OH 18-cm main lines, providing additional constraints on the evolution of diffuse molecular gas in the ISM \citep[][]{Gupta18oh, Balashev21}. 

In this paper, we present a spectroscopically blind search of \hi\ 21-cm absorption at $z>2$ based on a sample of AGNs selected using the mid-infrared (MIR) colors from  Wide-field Infrared Survey Explorer \citep[WISE;][]{Wright10, Cutri14} and having spectroscopically confirmed redshifts using the Southern African Large Telescope (SALT; 180 hrs) and the Nordic Optical Telescope \citep[NOT; 3 nights;][]{Krogager18}. 
Note that similar to the radio waveband the infrared wavelengths are also unaffected by dust obscuration.  These AGNs are being observed as part of MALS, which is a large project at the MeerKAT array in South Africa, to search \hi\ 21-cm and OH 18-cm lines at $z<2$.  The  upgraded Giant Metrewave Radio Telescope (uGMRT) survey presented here covers $2<z<5.1$.

%
The paper is laid out as follows.  
In Section~\ref{sec:samp}, we present the sample definition and its properties in the context of previous radio-selected samples to search for DLAs.
The details of uGMRT observations and data analysis to obtain the radio spectra and spectral line catalog are presented in Section~\ref{sec:obsdat}.
We provide the details of \hi\ 21-cm absorber detected from the survey in Section~\ref{sec:dets}.
In sections~\ref{sec:int} and \ref{sec:assoc}, we compute the incidences of intervening and associated \hi\ 21-cm absorption lines, respectively. 
In Section~\ref{sec:int}, we apply the same formalism to also derive the incidence of intervening OH absorption.
The availability of SALT-NOT spectra allows us to examine the properties of gas along the sight line using \lya\ and various metal absorption lines.  
In particular, for a subset of uGMRT targets ($z_e>2.7$) through deeper SALT observations we have discovered 6 DLAs and 1 candidate proximate DLA (PDLA i.e., DLA within 3000\,\kms\ of $z_q$).
In Section~\ref{sec:int}, we also present the properties of these \lya\ and metal line absorbers and discuss the nature of multi-phase ISM in the context of uGMRT survey results. 
The results and future prospects are summarized in Section~\ref{sec:summ}.

Throughout this paper we use the $\Lambda$CDM cosmology with $\Omega_m$=0.27, $\Omega_\Lambda$=0.73 and 
H$_{\rm o}$=71\,\kms\,Mpc$^{-1}$.

\section{Sample}      
\label{sec:samp}   

\subsection{Definition and properties}      
\label{sec:sampdef}   

\begin{figure} 
\centerline{\vbox{
\centerline{\hbox{ 
\includegraphics[trim = {0cm 0cm 0cm 0cm}, width=0.20\textwidth,angle=0]{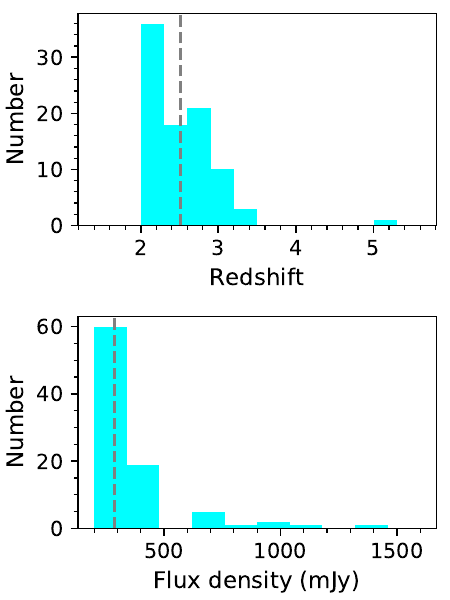}  
}} 
}}  
\vskip+0.0cm  
\caption{
Redshift and flux density (1.4\,GHz) distributions for our MIR selected sample.
The vertical dashed lines mark the median for each distribution.  
} 
\label{fig:samp}   
\end{figure} 

The targets for the uGMRT survey are drawn from the SALT-NOT sample of 303 AGNs constructed for MALS.  
The SALT-NOT sample is selected on the basis of MIR colors from WISE. We defined the following color wedge based on the first three bands of WISE i.e., $W1$ (3.4\,$\mu$m), $W2$ (4.6\,$\mu$m), $W3$ (12\,$\mu$m),   
\begin{equation}
\begin{array}{c}
W_1 - W_2 < 1.3\times(W_2 - W_3) - 3.04; \\
\\
W_1 - W_2 > 0.6.
\end{array}
\label{eqwedge}
\end{equation}
As shown in Fig.~1 of \citet[][]{Krogager18}, the MIR-wedge defined above is optimised  towards identifying most powerful AGNs (i.e., quasars) at $z>1.4$.  

The details of SALT-NOT target selection process will be presented in a future paper.  
In short, we cross-correlated AllWISE catalog \citep[][]{Cutri14} and radio sources brighter than $200$\,mJy in the NRAO VLA Sky Survey \citep[NVSS;][]{Condon98}, to identify 2011 high-probability quasar candidates satisfying the MIR wedge (Equation~\ref{eqwedge}).  We restricted the sample to declination $< +20^\circ$ to ensure reasonable observability with the MeerKAT telescope.
A search radius of $10^{\prime\prime}$ for WISE-NVSS cross-matching was used but all the coincidences were verified using higher spatial resolution quick look radio images at 3\,GHz from the Very Large Array Sky Survey \citep[VLASS; ][]{Lacy20}.  These quick look images have a spatial resolution of $\sim2.5^{\prime\prime}$ and the positional accuracy is limited to $\sim0.5^{\prime\prime}$.  
Consequently, our sample selects preferentially compact core-dominated AGNs.
We observed 303 candidates using SALT and NOT to measure redshifts and confirm the AGN nature. 
This optical spectroscopic campaign has led to a sample of AGNs which can be split into following three categories: {\it (i)} with emission lines in the optical spectrum (250 objects with confirmed redshifts at $0.1 < z < 5.1$), {\it (ii)} with no emission lines in the optical spectrum (26), and {\it (iii)} empty fields i.e., radio continuum peak coincides with the MIR source but neither emission line nor a continuum source are detected in optical spectra and images. 

The uGMRT {\tt Band-3} covers 250 - 500\,MHz which is suitable to search for \hi\ 21-cm absorption over $1.9 < z < 4.7$.  It nicely complements the MALS coverage of $z<1.4$. For the uGMRT survey presented here we selected all the 98 objects at $z>2$ from the SALT-NOT sample.  
In the allocated observing time we observed 88 of these which are listed in Table~\ref{tab:wisesamp}.  The redshift (median$\sim$2.5) and 1.4\,GHz flux density (median$\sim$288\,mJy) distributions are presented in Fig.~\ref{fig:samp}.  
The 1.4\,GHz spectral luminosities are in the range of $L_{\rm 1.4\,GHz} \simeq 10^{27 - 29.3}$\,W\,Hz$^{-1}$.  The lower end of luminosity is well above the radio cut-off that separates FRI and FRII radio sources, and the upper end corresponds to the most luminous radio-loud AGN at $z>5$ discovered from the SALT-NOT survey. 
All except one  are spectroscopically confirmed quasars. The details of radio galaxy M1540-1453 are presented by \citet[][]{Shukla21}.

With the right sample, it is possible to determine the evolution of cold gas in a dust-unbiased way.  Therefore, next we examine the efficacy of our sample selection strategy by comparing it with samples of DLAs from radio-selected quasars.

\subsection{Comparison with radio-selected DLA samples} 
\label{sec:sampcom}   

The three notable DLA samples based on radio-selected quasars are:
{\it (i)} the Complete Optical and Radio Absorption Line System (CORALS) survey of 66  QSOs ($z_{em} >2.2$) by \citet[][]{Ellison01},  {\it (ii)} the University of California San Diego (UCSD) survey of 53  QSOs ($z_{em} >2.0$) for DLAs by \citet[][]{Jorgenson06}, and 
{\it (iii)} the survey of 45 QSOs ($z_{em} >2.4$) selected from the Texas radio survey \citep[][]{Ellison08}. 
These surveys revealed 19, 7 and 9 DLAs over a 
redshift path, $\Delta z$ of 57.16, 41.15 and 38.79, respectively. 
The number of DLAs per unit redshift, $n_{\rm DLA}$, are estimated to be $0.31^{+0.09}_{-0.08}$,   $0.17^{+0.08}_{-0.07}$ and $0.23^{+0.11}_{-0.07}$, respectively.
The CORALS survey found a slightly higher incidence of DLAs and suggested that optically-selected DLA samples may be affected by dust-bias. 
But overall none of the surveys uncovered a population of dusty DLAs.  
%

\begin{figure} 
\centerline{\vbox{
\centerline{\hbox{ 
\includegraphics[trim = {0cm 0cm 0cm 0cm}, width=0.25\textwidth,angle=0]{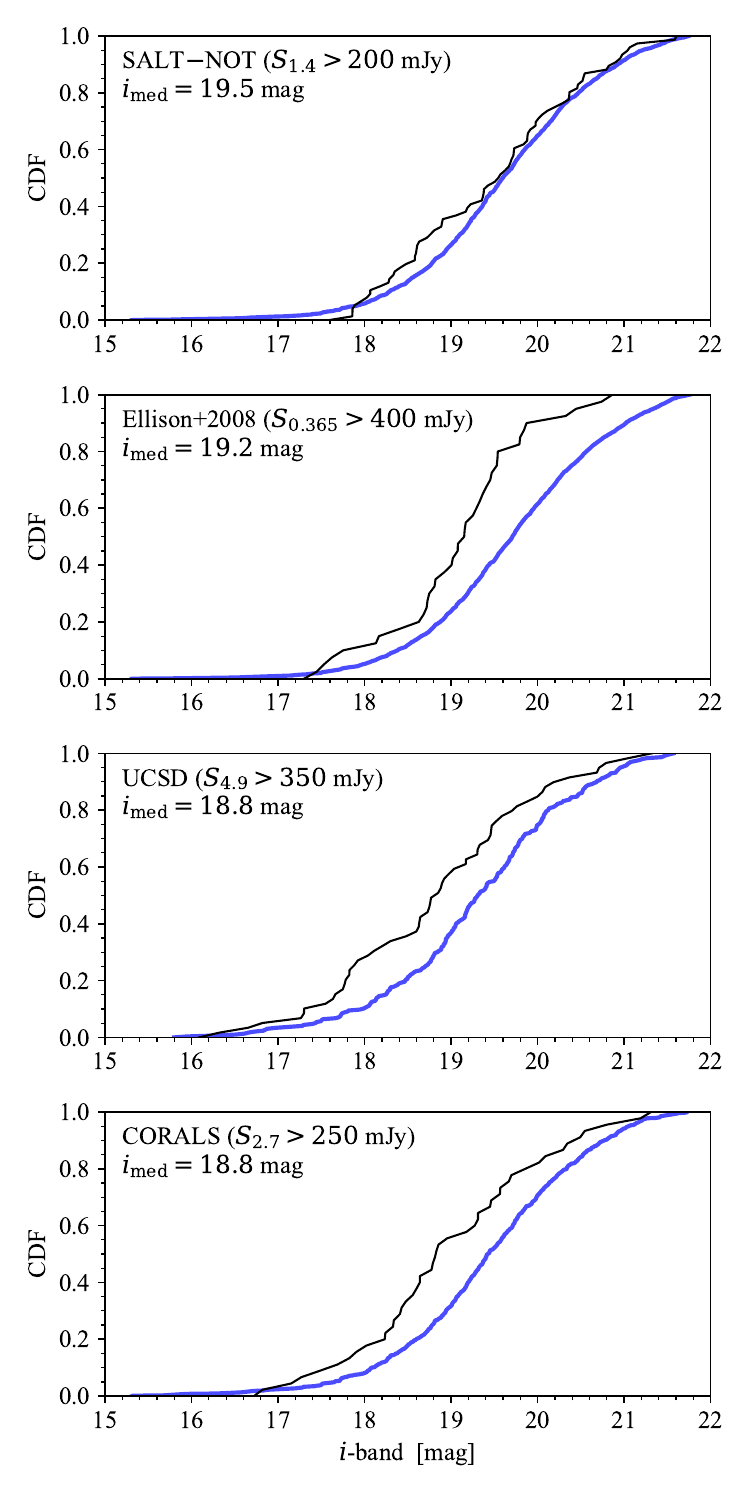}  
}} 
}}  
\vskip+0.0cm  
\caption{
Comparison of optical properties of the high-redshift ($z\gtrsim 2$), radio-selected quasar surveys: SALT-NOT $z>2$ (this work), \citet{Ellison08}, UCSD \citep{Jorgenson06}, and CORALS \citep{Ellison01}. The black line indicates the cumulative distribution of $i$-band magnitudes in the respective surveys, and the blue line shows the modelled distribution taking into account the survey radio flux limit and spectroscopic follow-up criterion of $B<22$~mag (see text). The median $i$-band magnitude of each sample is given in the upper left corner.
} 
\label{fig:comp}   
\end{figure} 

Targets for these three surveys have been selected at different radio frequencies and to different radio flux limits. While such differences might be subtle, they may still affect the optical properties of the quasars and hence the resulting statistics of DLAs. The CORALS survey has been selected at 2.7\,GHz down to a flux density limit of 250\,mJy; the UCSD sample has been selected at 4.9\,GHz to a flux limit of 350\,mJy; and lastly, the survey by \citet{Ellison08} has been selected at 356\,MHz down to a flux density limit of 400\,mJy.

In order to compare the effects of the radio-selection, we generate a mock distribution of $i$-band magnitudes for the three samples as well as for the SALT-NOT sample presented in this work. The intrinsic ultraviolet luminosity function is assumed to be the same in all cases and is taken from the work by \citet{Manti2017}. We assume a fixed distribution of the optical-to-radio flux ratio, $R_i$, following \citet{Balokovic2012} as well as a fixed radio slope of $\alpha_{\nu} = -0.8$, in order to scale the various survey limits to the same frequency (1.4\,GHz) as used in our survey and as used by \citet{Balokovic2012}. Since all the surveys impose roughly the same optical follow-up strategy in order to detect DLAs in low-resolution spectra, we impose a final cut on $B < 22$~mag. For this cut, we use an average color correction for high-redshift QSOs: $B = i + 0.3$ with a scatter of 0.1~mag \citep[see color relations by][]{Krogager19}. The resulting mock magnitude distribution is shown in Fig.~\ref{fig:comp} (blue curve) compared to the respective survey data (in black). While all surveys span a wide range of magnitudes, our survey more closely samples the underlying luminosity function and hence introduces a minimal bias in the optical properties of the sample.
This is a direct consequence of the fact that  SALT-NOT survey has targeted optically fainter quasars (refer to median $i$-band mags in Fig.~\ref{fig:comp}). 
An analysis of dust-bias in the sample using optical-infrared colors  will be presented in a future paper.  

\begin{longrotatetable}
\begin{deluxetable*}{lccclcccccccc}
	\tablecaption{Sample of $z>2$ MIR-selected radio sources (88) observed with uGMRT.}
\tabletypesize{\scriptsize}
\tablehead{
	\colhead{Source name} & \colhead{F$_{1.4\,GHz}$} & \colhead{$z_{em}$} & \colhead{Obs. run} & \colhead{Beam} & \colhead{F$_{\rm p,420\,MHz}$} & \colhead{F$_{\rm 420\,MHz}$ } & 
	\colhead{F$_{\rm p,420\,MHz}$/F$_{\rm 420\,MHz}$  } & \colhead{$\alpha^{1.4}_{0.4}$}  & \colhead{$\alpha_{inband}$} & \colhead{$\Delta$F} & \colhead{N$_{cand}$} \\ 
	\colhead{           } & \colhead{(mJy)} & \colhead{             }  & \colhead{       } & \colhead{       } & \colhead{ (mJy\,b$^{-1}$) } & \colhead{ (mJy)  } & 
	\colhead{    } & \colhead{  } & \colhead{  }  & \colhead{ (mJy\,b$^{-1}$) } \\ 
	\colhead{(1) } & \colhead{ (2) } & \colhead{ (3)     } & \colhead{ (4)} & \colhead{ (5) } & \colhead{  (6)    } & \colhead{ (7)       } & \colhead{  (8)  } &
	\colhead{(9)} & \colhead{ (10)}  & \colhead{ (11)} & \colhead{ (12)}  
} 
\startdata
M004243.06$+$124657.6        & 635.0     & 2.150   & 16SEP & $  7.0   ^{\prime\prime}\times   6.3  ^{\prime\prime},   -12.0 ^\circ $ & 1628.8  &  1882.5  &   0.87  &  $-$0.87 &  $-$0.86  &  1.9 &   -    \\ %
M005315.65$-$070233.4        & 248.2   & 2.130     & 16SEP & $  7.3   ^{\prime\prime}\times   6.5  ^{\prime\prime},   +17.0 ^\circ $ & 546.1   &  536.5   &   1.02  &  $-$0.62 &  $-$0.61  &  2.8 &   -    \\ %
M013047.38$-$172505.6        & 250.3   & 2.528     & 14SEP & $  9.9   ^{\prime\prime}\times   6.9  ^{\prime\prime},   +40.0 ^\circ $ & 532.4   &  560.2   &   0.95  &  $-$0.64 &  $-$0.66  &  1.5 &   -    \\ %
M021231.86$-$382256.6        & 244.5   & 2.260     & 14SEP & $  17.5  ^{\prime\prime}\times   6.8  ^{\prime\prime},   +33.0 ^\circ $ & 495.7   &  605.4   &   0.82  &  $-$0.72 &  $-$0.71  &  1.6 &   1    \\ %
M022613.72$+$093726.3        & 374.6   & 2.605     & 14SEP & $  8.9   ^{\prime\prime}\times   7.0  ^{\prime\prime},   +87.0 ^\circ $ & 435.1   &  436.6   &   1.0   &  $-$0.12 &  $-$0.02  &  2.1 &   -    \\ %
M022639.92$+$194110.1        & 209.8   & 2.190     & 14SEP & $  10.4  ^{\prime\prime}\times   6.7  ^{\prime\prime},   -83.0 ^\circ $ & 390.5   &  381.7   &   1.02  &  $-$0.48 &  $-$0.41  &  1.9 &   -    \\ %
M024939.93$+$044028.9        & 420.5   & 2.008     & 17SEP & $  9.1   ^{\prime\prime}\times   7.6  ^{\prime\prime},   +89.0 ^\circ $ & 927.7   &  992.7   &   0.93  &  $-$0.69 &  $-$0.67  &  1.6 &   -    \\ %
M025035.54$-$262743.1        & 389.2   & 2.918   & 17SEP & $  11.0  ^{\prime\prime}\times   6.7  ^{\prime\prime},   +25.0 ^\circ $ & 389.6   &  419.0   &   0.93  &  $-$0.06 &  $+$0.01  &  1.7 &   1    \\ %
M032808.59$-$015220.2        & 221.9   & 2.679   & 17SEP & $  12.0  ^{\prime\prime}\times   8.0  ^{\prime\prime},   +72.0 ^\circ $ & 370.2   &  527.3   &   0.70  &  $-$0.69 &  $-$0.63  &  1.3 &   -    \\ %
M041620.54$-$333931.3        & 264.1   & 3.045   & 08SEP & $  17.5  ^{\prime\prime}\times   9.0  ^{\prime\prime},   -42.0 ^\circ $ & 130.0   &  117.8   &   1.1   &  $+$0.64 &  $+$0.51  &  1.7 &   -    \\ %
M042248.53$-$203456.6        & 224.3   & 2.582   & 08SEP & $  11.8  ^{\prime\prime}\times   8.4  ^{\prime\prime},   -58.0 ^\circ $ & 187.3   &  192.1   &   0.98  &  $+$0.12 &  $+$0.16  &  1.7 &   1    \\ %
M044849.48$-$093531.3        & 240.9   & 2.079   & 14SEP & $  9.3   ^{\prime\prime}\times   7.2  ^{\prime\prime},   +46.0 ^\circ $ & 152.5   &  147.3   &   1.04  &  $+$0.39 &  $+$0.39  &  1.6 &   -    \\ %
M050725.04$-$362442.9        & 212.4   & 2.930   & 08SEP & $  15.9  ^{\prime\prime}\times   9.6  ^{\prime\prime},   -35.0 ^\circ $ & 494.5   &  474.5   &   1.04  &  $-$0.64 &  $-$0.51  &  1.6 &   -    \\ %
M051240.99$+$151723.8        & 966.5   & 2.568     & 07SEP & $  6.8   ^{\prime\prime}\times   6.5  ^{\prime\prime},   -79.0 ^\circ $ & 560.1   &  595.9   &   0.94  &  $+$0.39 &  $+$0.27  &  1.8 &   -    \\ %
M051340.03$+$010023.6        & 447.0   & 2.673   & 07SEP & $  7.7   ^{\prime\prime}\times   6.6  ^{\prime\prime},   +64.0 ^\circ $ & 342.5   &  349.7   &   0.98  &  $+$0.20 &  $-$0.04  &  2.1 &   4    \\ %
M051511.18$-$012002.4        & 288.8   & 2.287   & 07SEP & $  8.9   ^{\prime\prime}\times   6.4  ^{\prime\prime},   +59.0 ^\circ $ & 412.6   &  641.5   &   0.64  &  $-$0.64 &  $-$0.68  &  1.6 &   1    \\ %
M051656.35$+$073252.7        & 231.7   & 2.594     & 07SEP & $  10.6  ^{\prime\prime}\times   6.9  ^{\prime\prime},   +90.0 ^\circ $ & 44.2    &  44.1    &   1.0   &  $+$1.32 &  $+$1.10  &  1.9 &   -    \\ %
M052318.55$-$261409.6        & 1354.9  & 3.110   & 08SEP & $  12.0  ^{\prime\prime}\times   9.1  ^{\prime\prime},   -52.0 ^\circ $ & 477.7   &  451.8   &   1.06  &  $+$0.88 &  $+$1.08  &  2.2 &   -    \\ %
M061038.80$-$230145.6        & 360.2   & 2.829   & 14SEP & $  11.0  ^{\prime\prime}\times   7.0  ^{\prime\prime},   +31.0 ^\circ $ & 130.5   &  129.0   &   1.01  &  $+$0.82 &  $+$0.89  &  1.7 &   -    \\ %
M061856.02$-$315835.2        & 346.1   & 2.134   & 09SEP & $  10.8  ^{\prime\prime}\times   6.3  ^{\prime\prime},   +0.0  ^\circ $ & 877.1   &  828.6   &   1.06  &  $-$0.70 &  $-$0.35  &  2.0 &   1    \\ %
M063602.28$-$311312.5        & 262.1   & 2.654   & 09SEP & $  19.2  ^{\prime\prime}\times   11.1 ^{\prime\prime},   +33.0 ^\circ $ & 178.7   &  162.4   &   1.1   &  $+$0.38 &  $+$0.42  &  3.1 &   -    \\ %
M063613.53$-$310646.3        & 208.0   & 2.757   & 09SEP & $  11.6  ^{\prime\prime}\times   6.9  ^{\prime\prime},   +17.0 ^\circ $ & 436.2   &  474.2   &   0.92  &  $-$0.66 &  $-$0.73  &  2.1 &   -    \\ %
M065254.73$-$323022.6$^\dag$ & 322.1   & 2.239   & 08SEP & $  11.4  ^{\prime\prime}\times   10.1 ^{\prime\prime},   +5.0  ^\circ $ & 475.9   &  611.1   &   0.78  &  $-$0.85 &  $-$0.95  &  2.3 &   -    \\ %
                             &         &           &       &                                                                      & 279.0   &    328.2   &   0.85  &          &           &  2.3 &   -    \\ 
M070249.30$-$330205.0        & 314.6   & 2.410   & 08SEP & $  12.3  ^{\prime\prime}\times   9.8  ^{\prime\prime},   +33.0 ^\circ $ & 574.9   &  599.8   &   0.96  &  $-$0.52 &  $-$0.52  &  2.8 &   1    \\ 
M073159.01$+$143336.3        & 316.5   & 2.632     & 16SEP & $  8.8   ^{\prime\prime}\times   8.1  ^{\prime\prime},   -49.0 ^\circ $ & 180.0   &  185.4   &   0.97  &  $+$0.43 &  $+$0.40  &  7.8 &   -    \\ 
M073714.60$-$382841.9        & 219.3   & 2.107   & 14SEP & $  14.5  ^{\prime\prime}\times   6.6  ^{\prime\prime},   +18.0 ^\circ $ & 498.9   &  515.3   &   0.97  &  $-$0.68 &  $-$0.67  &  2.4 &   -    \\ 
M080804.34$+$005708.2        & 317.0   & 3.133   & 08SEP & $  14.3  ^{\prime\prime}\times   7.1  ^{\prime\prime},   -73.0 ^\circ $ & 434.8   &  450.7   &   0.96  &  $-$0.28 &  $-$0.26  &  2.1 &   -    \\ 
M081936.62$-$063047.9        & 280.0   & 2.507   & 08SEP & $  11.9  ^{\prime\prime}\times   6.7  ^{\prime\prime},   -73.0 ^\circ $ & 339.0   &  354.9   &   0.96  &  $-$0.19 &  $-$0.08  &  2.3 &   -    \\ 
M085826.92$-$260721.0        & 404.5   & 2.036   & 09SEP & $  17.3  ^{\prime\prime}\times   8.5  ^{\prime\prime},   -19.0 ^\circ $ & 561.0   &  523.9   &   1.07  &  $-$0.21 &  $-$0.28  &  2.6 &   -    \\ 
M090910.66$-$163753.8        & 340.1   & 2.475   & 09SEP & $  9.0   ^{\prime\prime}\times   7.3  ^{\prime\prime},   -11.0 ^\circ $ & 776.2   &  909.7   &   0.85  &  $-$0.79 &  $-$0.94  &  1.7 &   -    \\ 
M091051.01$-$052626.8$^\dag$ & 337.9   & 2.395   & 16SEP & $  9.6   ^{\prime\prime}\times   7.2  ^{\prime\prime},   -44.0 ^\circ $ & 151.4   &  166.3   &   0.91  &  $+$0.20 &  $+$0.29  &  1.6 &   -    \\ %
                             &         &           &       &                                                                         & 80.0    &  97.3    &   0.82  &          &           &  1.6 &   -    \\
M095231.66$-$245349.1        & 209.5   & 2.626   & 14SEP & $  10.4  ^{\prime\prime}\times   7.0  ^{\prime\prime},   +5.0  ^\circ $ & 224.2   &  210.3   &   1.07  &  $-$0.00 &  $-$0.17  &  1.8 &   -    \\ 
M100715.18$-$124746.7        & 381.1   & 2.113   & 16SEP & $  7.7   ^{\prime\prime}\times   6.2  ^{\prime\prime},   -24.0 ^\circ $ & 476.6   &  470.0   &   1.01  &  $-$0.17 &  $-$0.21  &  2.1 &   -    \\ 
M101313.10$-$254654.7        & 248.8   & 2.965   & 14SEP & $  10.6  ^{\prime\prime}\times   6.9  ^{\prime\prime},   +10.0 ^\circ $ & 216.7   &  218.6   &   0.99  &  $+$0.10 &  $-$0.0   &  2.1 &   -    \\ 
M102548.76$-$042933.0        & 363.5   & 2.292   & 16SEP & $  6.9   ^{\prime\prime}\times   6.7  ^{\prime\prime},   +89.0 ^\circ $ & 356.3   &  391.6   &   0.91  &  $-$0.06 &  $-$0.03  &  2.5 &   -    \\ 
M104314.53$-$232317.5        & 212.1   & 2.881   & 07SEP & $  8.9   ^{\prime\prime}\times   6.5  ^{\prime\prime},   -10.0 ^\circ $ & 460.6   &  505.8   &   0.91  &  $-$0.69 &  $-$0.79  &  2.2 &   -    \\ 
M111820.61$-$305459.0        & 233.2   & 2.352   & 08SEP & $  18.5  ^{\prime\prime}\times   8.4  ^{\prime\prime},   -51.0 ^\circ $ & 64.8    &  70.0    &   0.93  &  $+$0.96 &  $+$0.10  &  3.0 &   -    \\ 
M111917.36$-$052707.9        & 1174.4  & 2.651   & 07SEP & $  7.1   ^{\prime\prime}\times   6.3  ^{\prime\prime},   -48.0 ^\circ $ & 1696.6  &  1793.2  &   0.95  &  $-$0.34 &  $-$0.47  &  3.6 &   -    \\
M112402.56$-$150159.1        & 261.9   & 2.551   & 07SEP & $  9.0   ^{\prime\prime}\times   6.9  ^{\prime\prime},   -23.0 ^\circ $ & 194.7   &  196.0   &   0.99  &  $+$0.23 &  $+$0.01  &  2.6 &   -    \\ 
M114226.58$-$263313.7        & 294.7   & 3.237   & 08SEP & $  15.5  ^{\prime\prime}\times   8.3  ^{\prime\prime},   -56.0 ^\circ $ & 293.6   &  340.7   &   0.86  &  $-$0.35 &  $-$0.31  &  4.0 &   -    \\ 
M115222.04$-$270126.3        & 238.2   & 2.703   & 08SEP & $  13.3  ^{\prime\prime}\times   8.6  ^{\prime\prime},   -60.0 ^\circ $ & 410.4   &  441.4   &   0.93  &  $-$0.49 &  $-$0.64  &  2.5 &   -    \\ 
M115306.72$-$044254.5$^\dag$ & 684.8   & 2.591   & 16SEP & $  7.0   ^{\prime\prime}\times   6.5  ^{\prime\prime},   +86.0 ^\circ $ & 929.3   &  899.1   &   1.03  &  $-$0.78 &  $-$0.89  &  2.7 &   -    \\ 
                             &         &           &       &                                                                         & 980.1   &  912.2   &   1.07  &          &           &  2.7 &   -    \\ 
M120632.23$-$071452.6        & 698.8   & 2.263     & 16SEP & $  8.0   ^{\prime\prime}\times   6.2  ^{\prime\prime},   -9.0  ^\circ $ & 1402.8  &  1267.9  &   1.10  &  $-$0.48 &  $-$0.60  &  2.7 &   -    \\
M121514.42$-$062803.5        & 360.4   & 3.218   & 16SEP & $  9.2   ^{\prime\prime}\times   6.6  ^{\prime\prime},   -8.0  ^\circ $ & 511.1   &  461.5   &   1.11  &  $-$0.20 &  $-$0.31  &  2.8 &   -    \\ 
M123150.30$-$123637.5        & 276.0   & 2.106     & 07SEP & $  7.2   ^{\prime\prime}\times   6.5  ^{\prime\prime},   -17.0 ^\circ $ & 159.7   &  205.1   &   0.78  &  $+$0.24 &  $+$0.36  &  1.8 &   -    \\ 
M123410.08$-$332638.5        & 297.9   & 2.820   & 08SEP & $  15.8  ^{\prime\prime}\times   9.7  ^{\prime\prime},   -57.0 ^\circ $ & 665.4   &  710.4   &   0.94  &  $-$0.69 &  $-$0.72  &  3.4 &   -    \\ 
M124448.99$-$044610.2        & 384.9   & 3.104   & 16SEP & $  9.1   ^{\prime\prime}\times   7.7  ^{\prime\prime},   +28.0 ^\circ $ & 677.2   &  616.1   &   1.1   &  $-$0.38 &  $-$0.40  &  2.3 &   1    \\ 
M125442.98$-$383356.4        & 219.2   & 2.776   & 17SEP & $  14.7  ^{\prime\prime}\times   7.0  ^{\prime\prime},   +7.0  ^\circ $ & 297.4   &  300.9   &   0.99  &  $-$0.25 &  $-$0.25  &  2.3 &   -    \\
M125611.49$-$214411.7        & 260.7   & 2.178   & 16SEP & $  10.3  ^{\prime\prime}\times   6.5  ^{\prime\prime},   +13.0 ^\circ $ & 446.8   &  455.4   &   0.98  &  $-$0.45 &  $-$0.15  &  1.7 &   -    \\ 
M131207.86$-$202652.4        & 778.1   & 5.064   & 16SEP & $  10.5  ^{\prime\prime}\times   6.7  ^{\prime\prime},   +21.0 ^\circ $ & 1727.9  &  1721.9  &   1.0   &  $-$0.63 &  $-$0.62  &  2.0 &   1    \\ 
                             &         &           & 21APR$^\ddag$ & $  15.8  ^{\prime\prime}\times   13.8  ^{\prime\prime},   -38.0 ^{\circ \ddag}$ & -  &  2577.0$^\ddag$  &   -   &  - &  -  &  - &   -    \\ 
M132657.20$-$280831.4        & 404.5   & 2.238   & 16SEP & $  15.0  ^{\prime\prime}\times   7.0  ^{\prime\prime},   +20.0 ^\circ $ & 671.3   &  874.8   &   0.77  &  $-$0.62 &  $-$0.83  &  2.7 &   -    \\ 
M135131.98$-$101932.9        & 726.1   & 2.999   & 16SEP & $  9.2   ^{\prime\prime}\times   7.0  ^{\prime\prime},   +47.0 ^\circ $ & 1338.3  &  1886.5  &   0.71  &  $-$0.76 &  $-$0.79  &  2.0 &   -    \\ 
M141327.20$-$342235.1        & 274.7   & 2.812   & 09SEP & $  20.9  ^{\prime\prime}\times   6.7  ^{\prime\prime},   -42.0 ^\circ $ & 78.2    &  76.4    &   1.02  &  $+$1.02 &  $+$0.81  &  3.9 &   -    \\ 
M143709.04$-$294718.5        & 273.8   & 2.331   & 09SEP & $  14.6  ^{\prime\prime}\times   7.0  ^{\prime\prime},   -44.0 ^\circ $ & 333.0   &  456.1   &   0.73  &  $-$0.41 &  $-$0.49  &  2.5 &   -    \\ 
M144851.10$-$112215.6        & 455.5   & 2.630   & 07SEP & $  7.3   ^{\prime\prime}\times   6.0  ^{\prime\prime},   -48.0 ^\circ $ & 967.0   &  896.8   &   1.08  &  $-$0.54 &  $-$0.71  &  3.1 &   1    \\ 
M145342.95$-$132735.2        & 254.5   & 2.370   & 07SEP & $  7.6   ^{\prime\prime}\times   6.2  ^{\prime\prime},   -33.0 ^\circ $ & 477.3   &  634.7   &   0.75  &  $-$0.73 &  $-$0.83  &  2.2 &   -    \\ 
M145502.84$-$170014.2        & 294.7   & 2.291   & 07SEP & $  8.3   ^{\prime\prime}\times   6.4  ^{\prime\prime},   -13.0 ^\circ $ & 345.8   &  352.2   &   0.98  &  $-$0.14 &  $-$0.23  &  2.3 &   -    \\
M145625.83$+$045645.2        & 287.9   & 2.134   & 09SEP & $  7.1   ^{\prime\prime}\times   6.8  ^{\prime\prime},   +87.0 ^\circ $ & 800.8   &  813.2   &   0.98  &  $-$0.83 &  $-$0.82  &  2.6 &   -    \\ 
M145908.92$-$164542.3        & 378.9   & 2.006   & 07SEP & $  8.5   ^{\prime\prime}\times   6.6  ^{\prime\prime},   -9.0  ^\circ $ & 853.1   &  909.6   &   0.94  &  $-$0.70 &  $-$0.85  &  2.1 &   -    \\
M150425.30$+$081858.6        & 210.8   & 2.035   & 09SEP & $  8.4   ^{\prime\prime}\times   7.6  ^{\prime\prime},   -32.0 ^\circ $ & 122.1   &  138.5   &   0.88  &  $+$0.34 &  $+$0.27  &  4.8 &   -    \\ 
M151129.01$-$072255.3        & 326.3   & 2.582   & 17SEP & $  7.8   ^{\prime\prime}\times   6.8  ^{\prime\prime},   -85.0 ^\circ $ & 672.7   &  624.9   &   1.08  &  $-$0.52 &  $-$0.71  &  2.7 &   -    \\ 
M151304.72$-$252439.7$^\dag$ & 217.6   & 3.132   & 09SEP & $  9.1   ^{\prime\prime}\times   6.8  ^{\prime\prime},   +1.0  ^\circ $ & 855.7   &  819.4   &   1.04  &  $-$1.26 &  $-$1.26  &  3.6 &   -    \\
                             &         &           &       &                                                                         & 268.5   &  242.1   &   1.11  &          &           &  3.6 &   -    \\ 
M151944.77$-$115144.6        & 441.0   & 2.014   & 17SEP & $  9.2   ^{\prime\prime}\times   7.6  ^{\prime\prime},   +84.0 ^\circ $ & 425.0   &  561.5   &   0.76  &  $-$0.19 &  $-$0.01  &  3.7 &   -    \\ 
M154015.23$-$145341.5        & 203.3   & 2.098   & 17SEP & $  9.8   ^{\prime\prime}\times   7.5  ^{\prime\prime},   +61.0 ^\circ $ & 642.4   &  595.0   &   1.08  &  $-$0.86 &  $-$0.98  &  2.6 &   1    \\ 
M155825.35$-$215511.1        & 206.9   & 2.760   & 09SEP & $  13.7  ^{\prime\prime}\times   6.7  ^{\prime\prime},   -42.0 ^\circ $ & 209.5   &  235.8   &   0.89  &  $-$0.10 &  $+$0.09  &  3.0 &   -    \\ 
M161907.44$-$093952.5        & 340.3   & 2.891   & 17SEP & $  8.9   ^{\prime\prime}\times   6.8  ^{\prime\prime},   +66.0 ^\circ $ & 757.0   &  695.5   &   1.09  &  $-$0.57 &  $-$0.43  &  2.2 &   -    \\
M162047.94$+$003653.2        & 317.8   & 2.438   & 17SEP & $  10.7  ^{\prime\prime}\times   7.6  ^{\prime\prime},   +2.0  ^\circ $ & 226.9   &  234.3   &   0.97  &  $+$0.24 &  $+$0.16  &  3.3 &   -    \\ 
M164950.51$+$062653.3        & 389.2   & 2.144   & 17SEP & $  13.6  ^{\prime\prime}\times   9.3  ^{\prime\prime},   -11.0 ^\circ $ & 154.3   &  204.3   &   0.76  &  $+$0.51 &  $+$0.36  &  2.6 &   -    \\ 
M165038.03$-$124854.5        & 275.5   & 2.527   & 09SEP & $  7.3   ^{\prime\prime}\times   6.2  ^{\prime\prime},   -20.0 ^\circ $ & 729.6   &  675.1   &   1.08  &  $-$0.72 &  $-$0.52  &  5.0 &   -    \\ 
M165435.38$+$001719.2        & 255.3   & 2.363   & 07SEP & $  9.2   ^{\prime\prime}\times   7.3  ^{\prime\prime},   -14.0 ^\circ $ & 405.7   &  381.5   &   1.06  &  $-$0.32 &  $-$0.44  &  4.2 &   -    \\ 
M194110.28$-$300720.9        & 315.0   & 2.059   & 17SEP & $  23.6  ^{\prime\prime}\times   8.3  ^{\prime\prime},   +51.0 ^\circ $ & 164.0   &  227.2   &   0.72  &  $+$0.26 &  $+$0.40  &  3.6 &   -    \\
M200209.37$-$145531.8        & 620.3   & 2.192   & 17SEP & $  8.6   ^{\prime\prime}\times   7.1  ^{\prime\prime},   -17.0 ^\circ $ & 942.0   &  896.1   &   1.05  &  $-$0.29 &  $-$0.31  &  2.9 &   -    \\ 
M201708.96$-$293354.7        & 327.2   & 2.617   & 17SEP & $  20.5  ^{\prime\prime}\times   7.3  ^{\prime\prime},   +50.0 ^\circ $ & 1044.7  &  1201.1  &   0.87  &  $-$1.04 &  $-$1.02  &  2.7 &   1    \\ 
M203425.65$-$052332.2        & 419.7   & 2.070     & 17SEP & $  8.0   ^{\prime\prime}\times   7.3  ^{\prime\prime},   +85.0 ^\circ $ & 366.2   &  407.2   &   0.9   &  $+$0.02 &  $+$0.10  &  4.0 &   -    \\ 
M204737.67$-$184141.2        & 241.7   & 2.994   & 16SEP & $  13.8  ^{\prime\prime}\times   6.7  ^{\prime\prime},   -49.0 ^\circ $ & 273.0   &  284.3   &   0.96  &  $-$0.13 &  $-$0.17  &  2.2 &   -    \\ 
M205245.03$-$223410.6        & 330.9   & 2.072   & 16SEP & $  12.3  ^{\prime\prime}\times   6.0  ^{\prime\prime},   -41.0 ^\circ $ & 631.8   &  608.3   &   1.04  &  $-$0.49 &  $-$0.60  &  3.5 &   -    \\ 
M210143.29$-$174759.2        & 959.5   & 2.803   & 16SEP & $  9.5   ^{\prime\prime}\times   5.8  ^{\prime\prime},   -39.0 ^\circ $ & 2554.1  &  2477.5  &   1.03  &  $-$0.76 &  $-$0.91  &  3.6 &   -    \\
M212821.83$-$150453.2        & 245.5   & 2.547   & 17SEP & $  9.2   ^{\prime\prime}\times   6.9  ^{\prime\prime},   +6.0  ^\circ $ & 443.4   &  460.8   &   0.96  &  $-$0.50 &  $-$0.49  &  2.0 &   -    \\ 
M220127.50$+$031215.6        & 300.5   & 2.181   & 17SEP & $  18.3  ^{\prime\prime}\times   8.6  ^{\prime\prime},   +77.0 ^\circ $ & 180.5   &  191.6   &   0.94  &  $+$0.36 &  $+$0.33  &  1.7 &   -    \\ 
M222332.81$-$310117.3        & 231.7   & 3.206   & 17SEP & $  25.0  ^{\prime\prime}\times   9.2  ^{\prime\prime},   +40.0 ^\circ $ & 230.6   &  330.2   &   0.7   &  $-$0.28 &  $-$0.40  &  2.6 &   -    \\ 
M223816.27$-$124036.4        & 213.6   & 2.623   & 16SEP & $  12.3  ^{\prime\prime}\times   8.3  ^{\prime\prime},   -50.0 ^\circ $ & 464.3   &  435.5   &   1.07  &  $-$0.57 &  $-$0.65  &  3.7 &   -    \\ 
M224111.48$-$244239.0        & 211.4   & 2.242   & 16SEP & $  10.9  ^{\prime\prime}\times   6.1  ^{\prime\prime},   -31.0 ^\circ $ & 253.0   &  254.8   &   0.99  &  $-$0.15 &  $-$0.17  &  2.2 &   -    \\ 
M224705.52$+$121151.4        & 223.7   & 2.185     & 14SEP & $  12.1  ^{\prime\prime}\times   7.5  ^{\prime\prime},   +83.0 ^\circ $ & 474.3   &  489.1   &   0.97  &  $-$0.62 &  $-$0.60  &  1.9 &   -    \\ 
M224950.57$-$263459.6        & 228.8   & 2.174   & 16SEP & $  9.8   ^{\prime\prime}\times   5.7  ^{\prime\prime},   -23.0 ^\circ $ & 568.3   &  542.1   &   1.05  &  $-$0.69 &  $-$0.59  &  2.4 &   -    \\ 
M230036.41$+$194002.9        & 210.4   & 2.160     & 17SEP & $  27.7  ^{\prime\prime}\times   8.2  ^{\prime\prime},   +79.0 ^\circ $ & 475.8   &  556.0   &   0.86  &  $-$0.78 &  $-$0.65  &  11.2&   -    \\ 
M231634.61$+$042940.2        & 214.0   & 2.180   & 16SEP & $  6.9   ^{\prime\prime}\times   6.6  ^{\prime\prime},   +47.0 ^\circ $ & 103.3   &  97.9    &   1.06  &  $+$0.70 &  $+$0.50  &  3.0 &   -    \\ 
M234910.12$-$043803.2        & 206.1   & 2.240   & 16SEP & $  7.7   ^{\prime\prime}\times   6.8  ^{\prime\prime},   -62.0 ^\circ $ & 168.3   &  185.6   &   0.91  &  $+$0.08 &  $-$0.04  &  5.1 &   -    \\
M235722.47$-$073134.3        & 235.5   & 2.764   & 16SEP & $  7.0   ^{\prime\prime}\times   6.0  ^{\prime\prime},   -19.0 ^\circ $ & 372.4   &  398.2   &   0.94  &  $-$0.42 &  $-$0.40  &  2.4 &   -    \\ 
\label{tab:wisesamp}
\enddata
\tablecomments{ 
Column 1: source name based on right ascension and declination (J2000) from NVSS. Column 2: 20\,cm flux density from NVSS. Column 3: emission line redshift measured from SALT-NOT survey. Column 4: observing run (see Table~\ref{tab:obslog}). Note that only M1312-2026 was also observed at {\tt Band-2}. 
Columns 5 - 7: synthesised beam, peak of the prominent Gaussian component and total flux densities, respectively, from the continuum image based on 390-450\,MHz range (average 420\,MHz). Column 8: ratio of columns 6 and 7.  Column 9: spectral index derived using NVSS and 420\,MHz flux densities. In a few cases, which are all sources with a single component fit, this ratio marginally ($\sim$5\%) exceeds 1, suggesting that the radio emission may be partially resolved.
Column 10: in-band spectral index.
Column 11: observed spectral rms at 420\,MHz. Column 12: number of absorption candidates. \\
$\dag$: The radio source is double lobed in 420\,MHz image (see Section~\ref{sec:obs} for details). $\ddag$: Corresponds to {\tt Band-2}.
}
\end{deluxetable*}
\end{longrotatetable}

\section{Observations and data analysis}      
\label{sec:obsdat}   

\subsection{Observations}      
\label{sec:obs}   

We used the recently commissioned {\tt Band-2} (120-240\,MHz) and {\tt Band-3} (250-500\,MHz) of uGMRT to observe redshifted associated and intervening \hi\ 21-cm absorption lines from the sample. The total allocated time including 
all overheads for the survey observations was 90\,hrs.  
%

The {\tt Band-3} observations were split into 6 observing runs in September, 2018 (see Table~\ref{tab:obslog}).  
For these, we used GMRT Wideband Backend (GWB) with a baseband bandwidth of 200\,MHz covering 300-500\,MHz and split into 8192 frequency channels.  This corresponds to a redshift coverage of 1.84 - 3.73 for \hi\ 21-cm line.  The channel resolution is 24.414\,kHz, which at 400\,MHz provides a velocity resolution of 18.3\,\kms. 
Each target was observed for typically 30-45\,mins. 
The details of which target sources were visited in which observing run are summarized in column\,4 of Table~\ref{tab:wisesamp}.  

For {\tt Band-2} observations which targeted only M1312-2026, the highest redshift quasar in the sample, the GMRT Software Backend (GSB) was used to configure a baseband bandwidth of 4.17\,MHz split into 512 spectral channels. The observing band was centered at 234.1\,MHz (resolution$\sim$10\,\kms), the redshifted \hi\ 21-cm line frequency of the source. The total on-source time was 4.2\,hrs. 

Additionally, five absorption candidates identified from the {\tt Band-3} survey observations were reobserved on December 10, 2019 and February 20, 2020.  We used GWB with a bandwidth of 6.25\,MHz centered at line frequency (details in Section~\ref{sec:specan}) and split into 4096 channels (resolution$\sim$1\,\kms). 
Each candidate was observed for 3\,hrs (on-source time $\sim$2.2\,hrs). 

For all the observations, only parallel hand correlations RR and LL were obtained.  During each observing run 3C48, 3C147 and/or 3C286 were observed for flux density and bandpass calibrations.  A complex gain calibrator was also observed for each target source.

\begin{deluxetable}{cccc}
\tablecaption{Details of uGMRT observations.}
\tablehead{
\colhead{ Run ID} &  \colhead{Band     } & \colhead{Date$^\dag$} & \colhead{Duration$^\ddag$} \\
}
\startdata
\vspace{-0.1cm} \\
\multicolumn{4}{c}{\small {\bf Survey observations}}   \\
21APR           &  {\tt Band-2}     & 2018 April 21         &  7  \\
07SEP           &  {\tt Band-3}     & 2018 September 07     &  11 \\
08SEP           &      ''           & 2018 September 08     &  10 \\
09SEP           &      ''           & 2018 September 09     &  11 \\
14SEP           &      ''           & 2018 September 14     &  10 \\
16SEP           &      ''           & 2018 September 16     &  21 \\
17SEP           &      ''           & 2018 September 17     &  20 \\
\vspace{-0.1cm} \\
\multicolumn{4}{c}{\small {\bf Follow-up observations}}   \\
10DEC           &      ''           & 2019 December 10      &  6  \\
20FEB           &      ''           & 2020 February 20      &  9  \\
\label{tab:obslog}
\enddata
\tablecomments{ 
$\dag$: Start date as per Indian Standard Time. $\ddag$: In hours.
}
\end{deluxetable}

\begin{figure*} 
\centerline{\vbox{
\centerline{\hbox{ 
\includegraphics[trim = {0cm 0cm 0cm 0cm}, width=0.30\textwidth,angle=0]{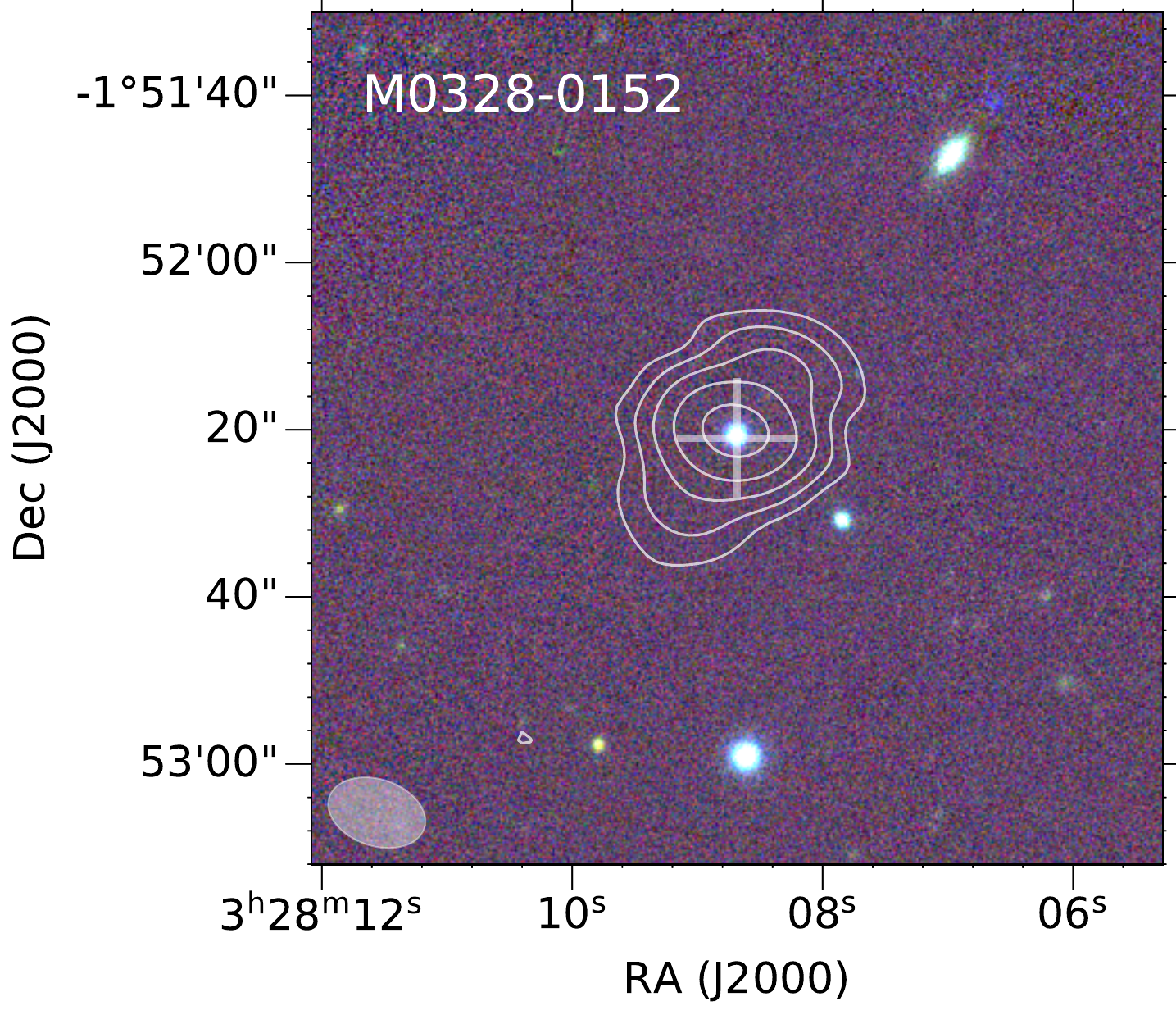} 
\includegraphics[trim = {0cm 0cm 0cm 0cm}, width=0.31\textwidth,angle=0]{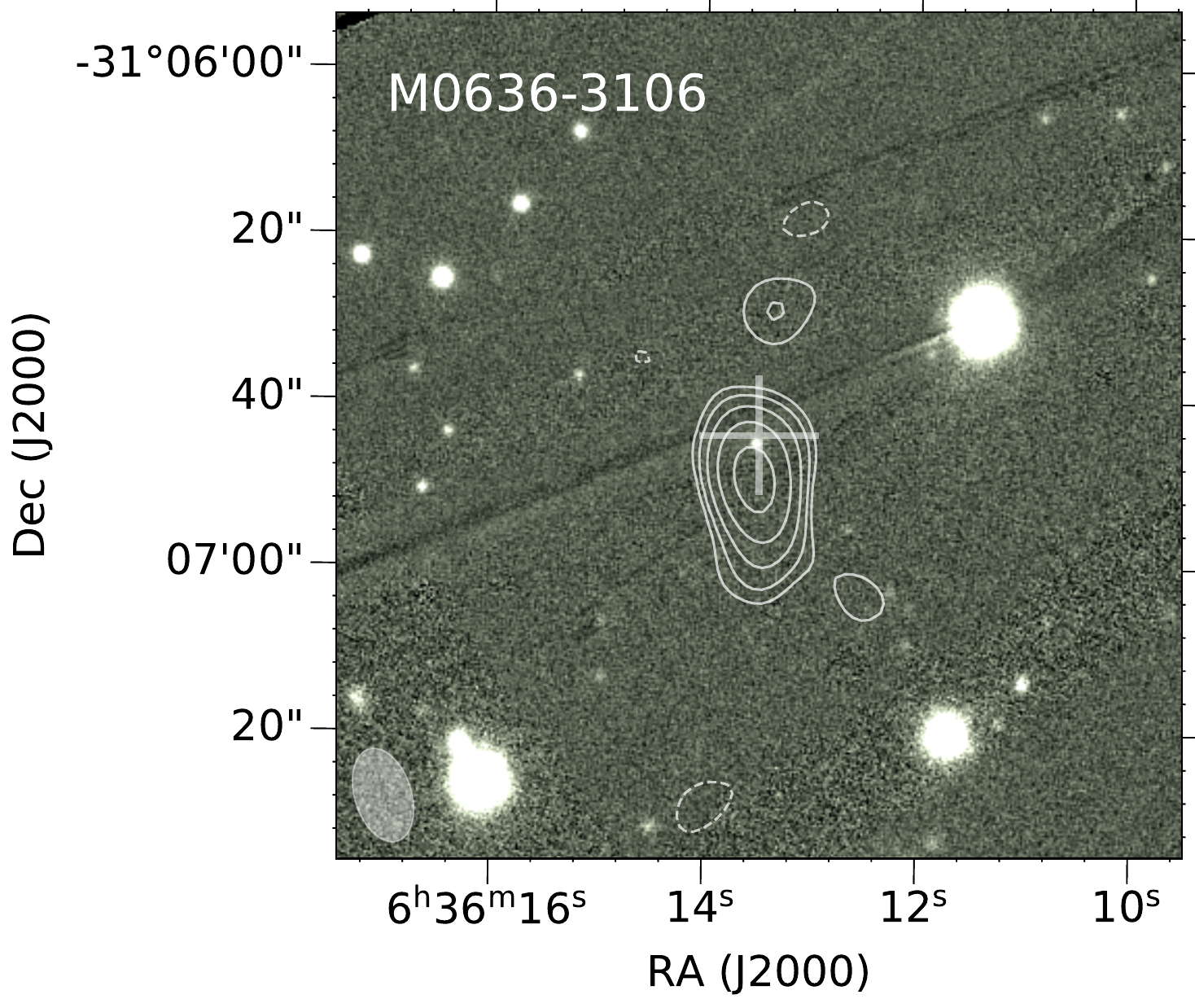} 
\includegraphics[trim = {0cm 0cm 0cm 0cm}, width=0.315\textwidth,angle=0]{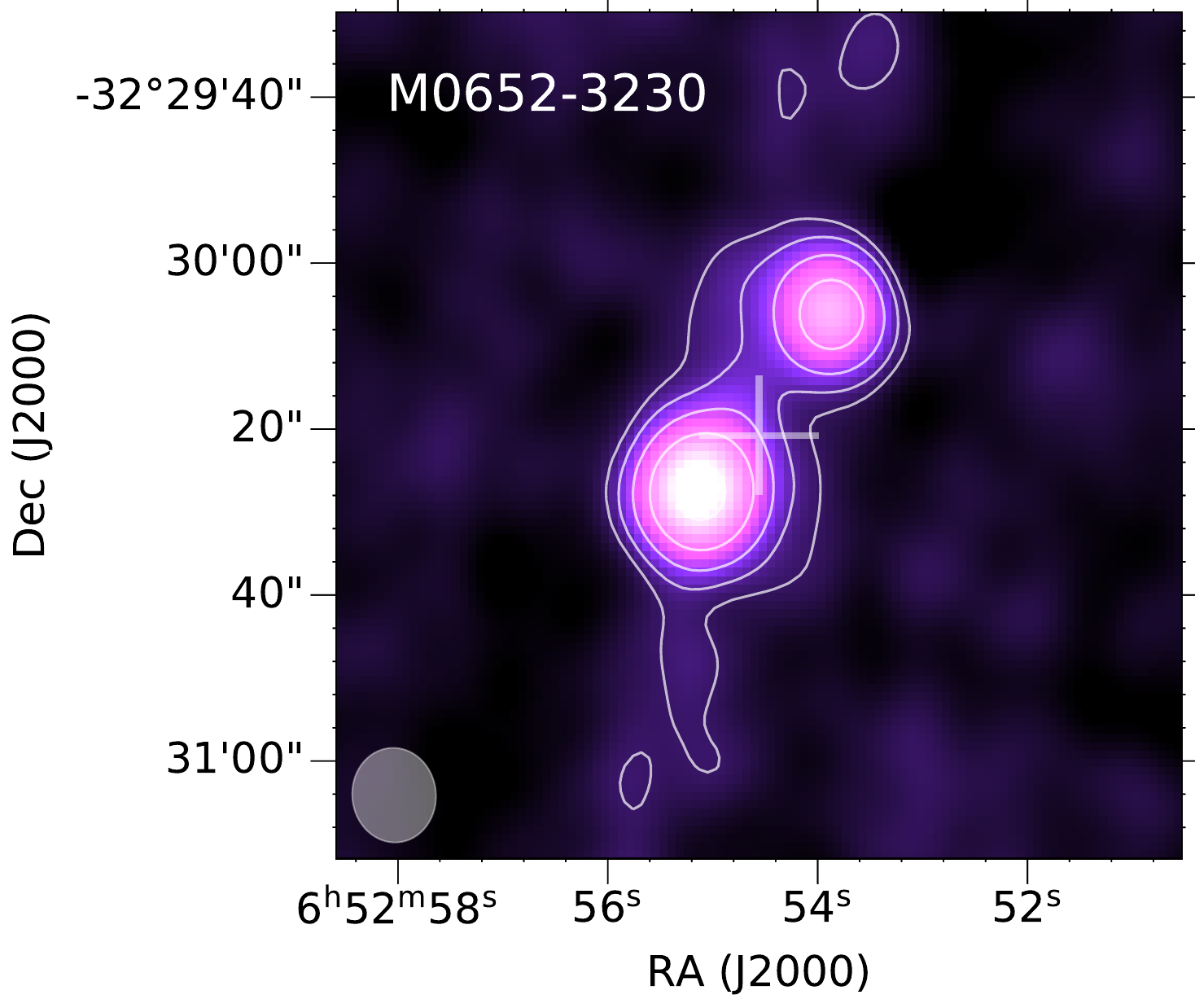} 
}}
\centerline{\hbox{ 
\includegraphics[trim = {0cm 0cm 0cm 0cm}, width=0.30\textwidth,angle=0]{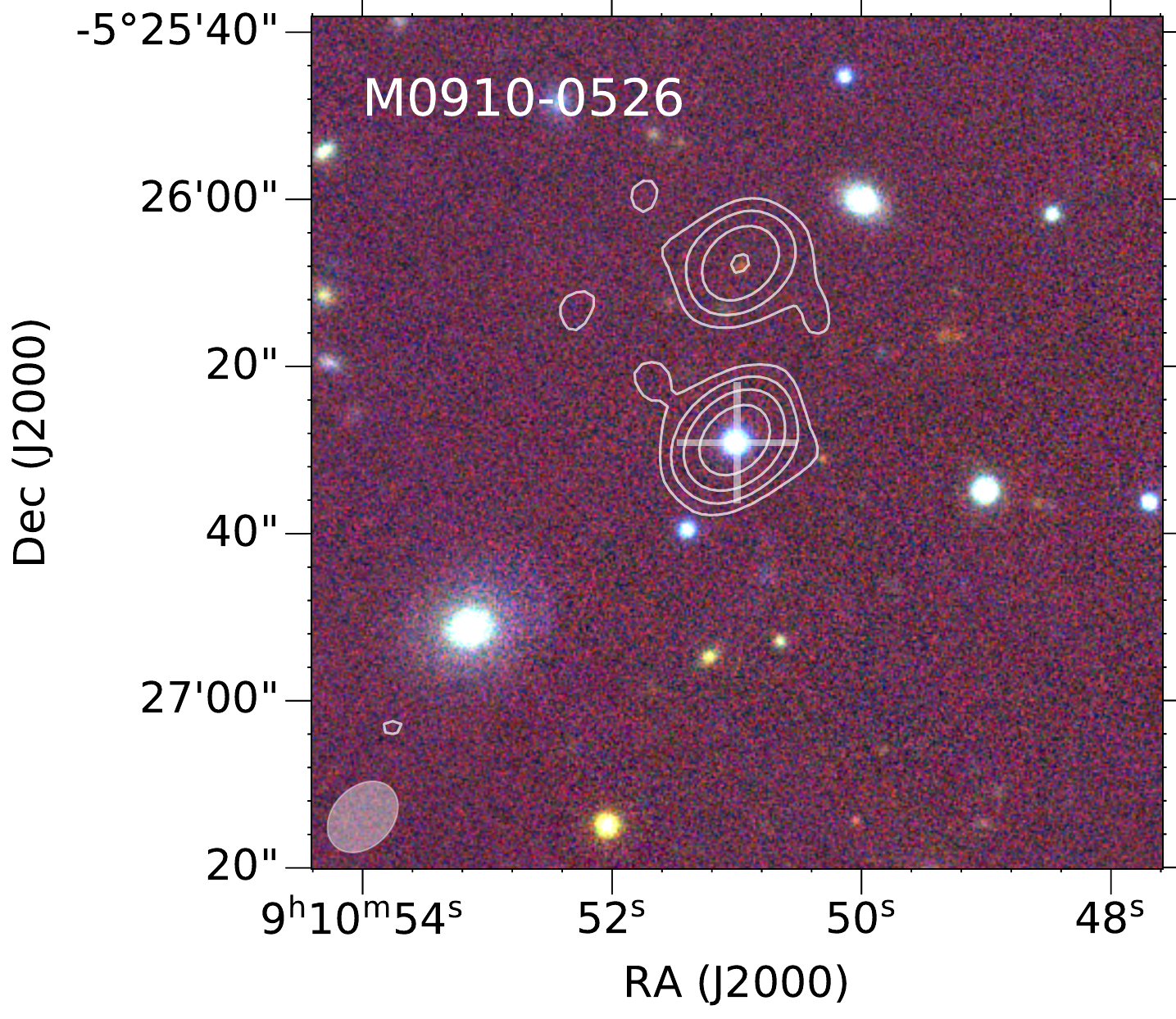}  
\includegraphics[trim = {0cm 0cm 0cm 0cm}, width=0.31\textwidth,angle=0]{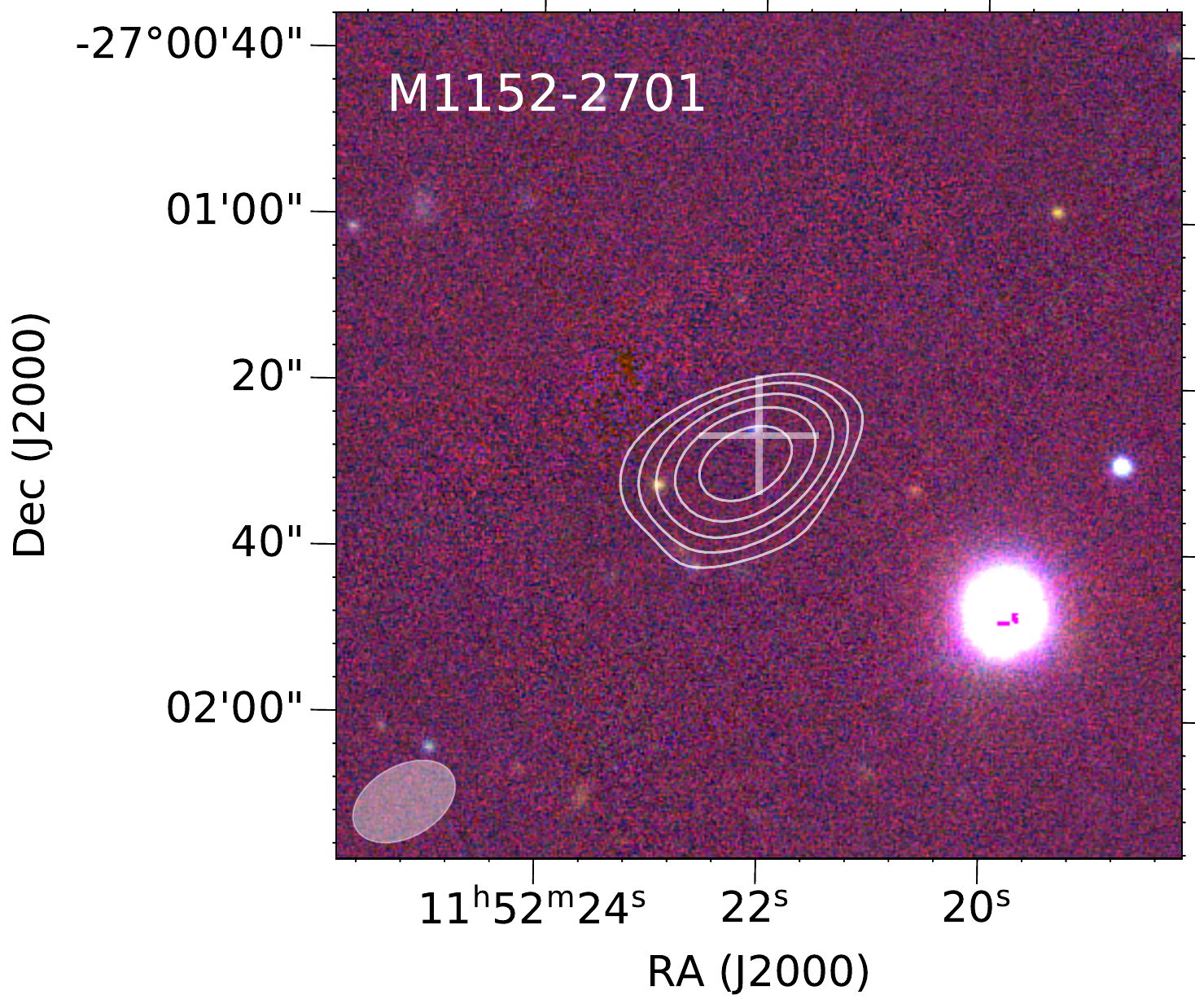}  
\includegraphics[trim = {0cm 0cm 0cm 0cm}, width=0.30\textwidth,angle=0]{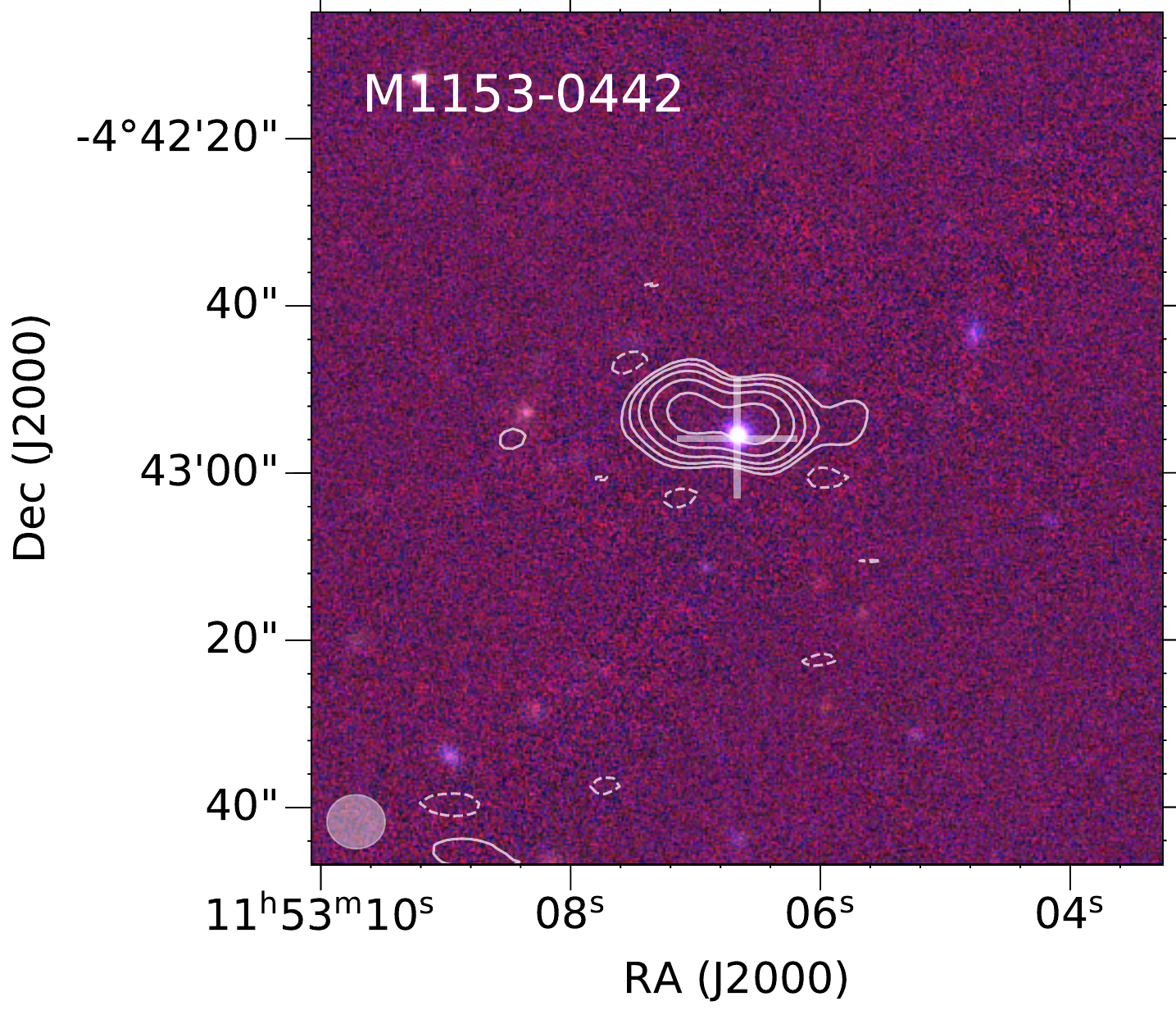} 
}}
\centerline{\hbox{ 
\includegraphics[trim = {0cm 0cm 0cm 0cm}, width=0.30\textwidth,angle=0]{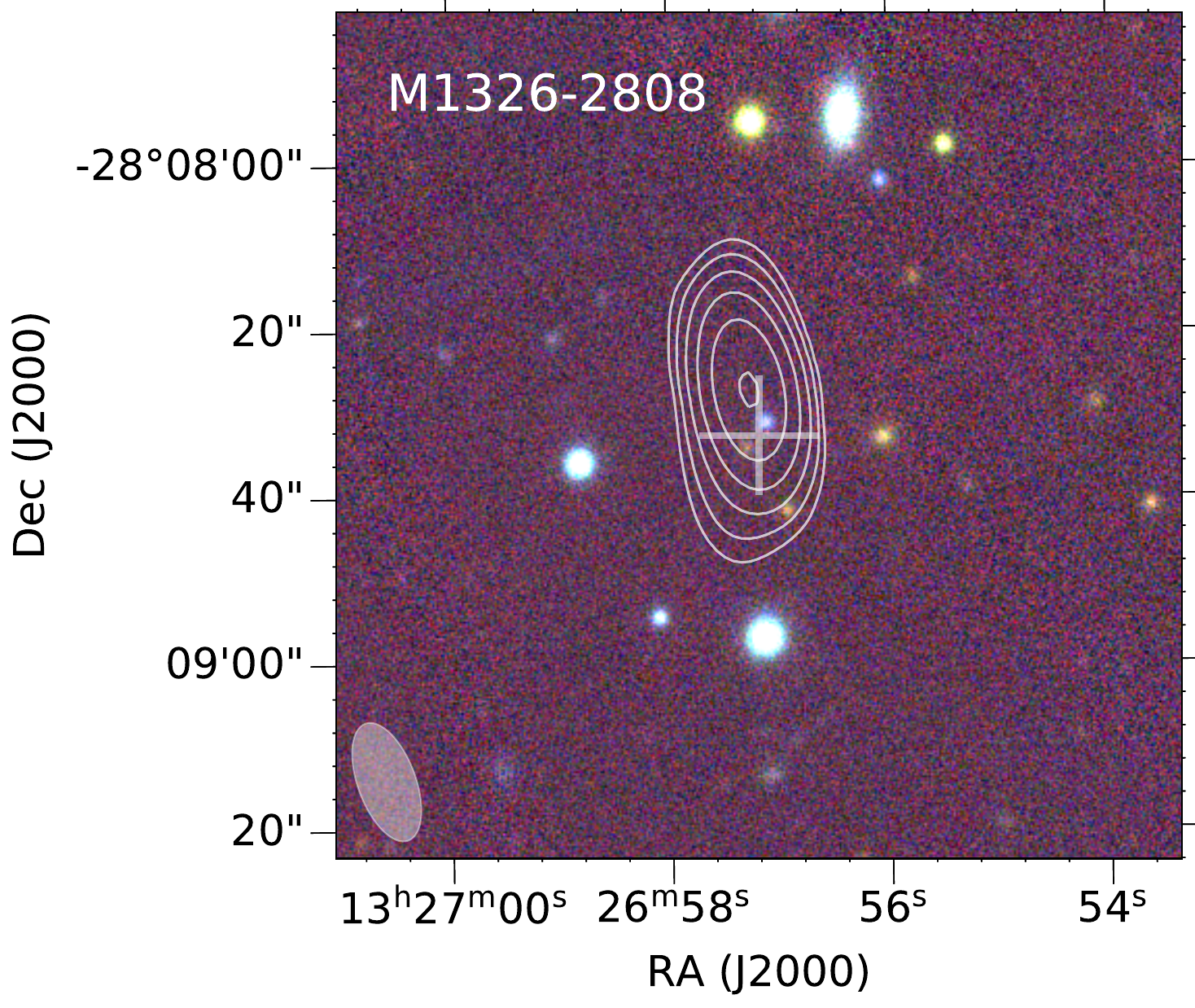}  
\includegraphics[trim = {0cm 0cm 0cm 0cm}, width=0.30\textwidth,angle=0]{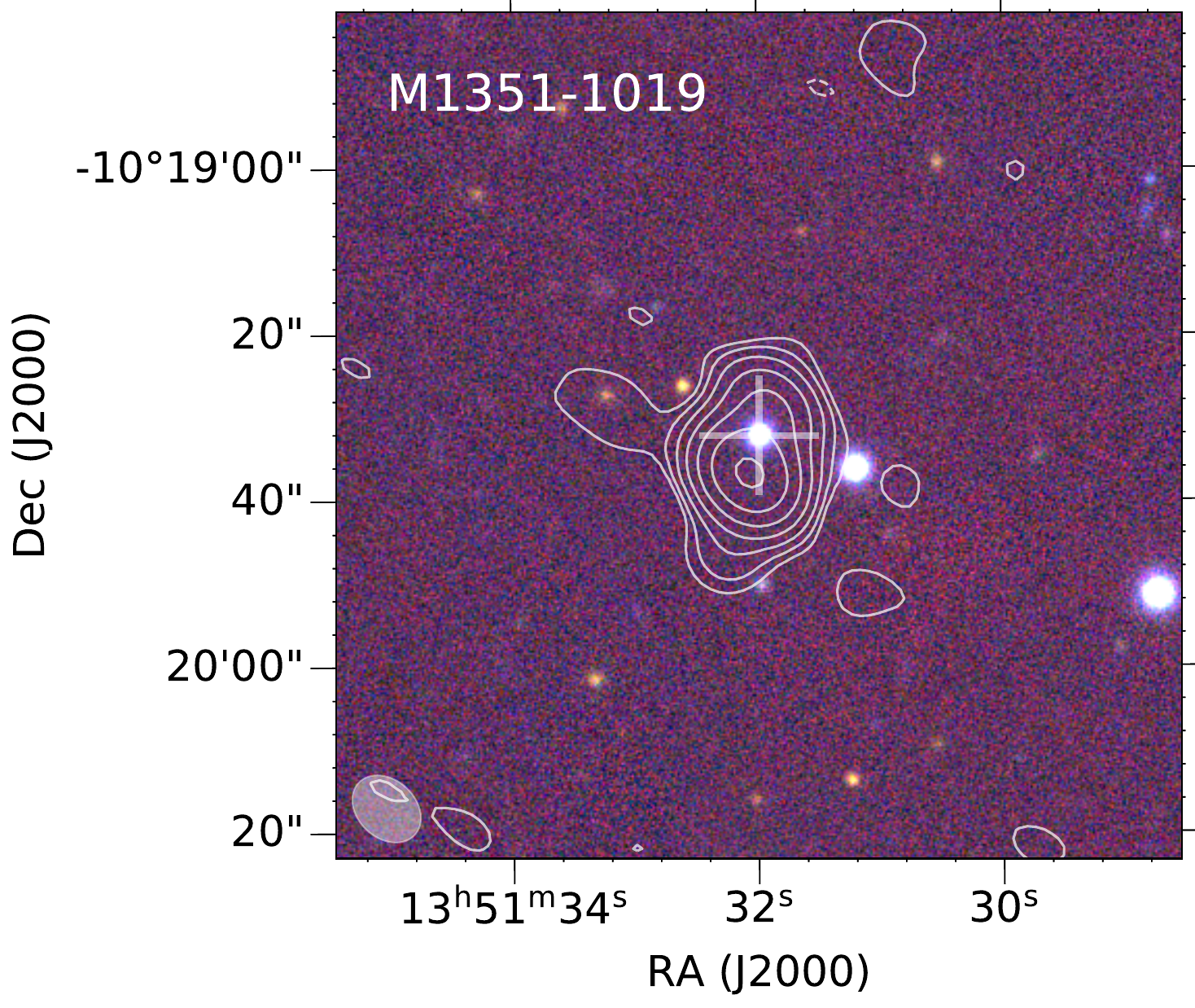}  
\includegraphics[trim = {0cm 0cm 0cm 0cm}, width=0.30\textwidth,angle=0]{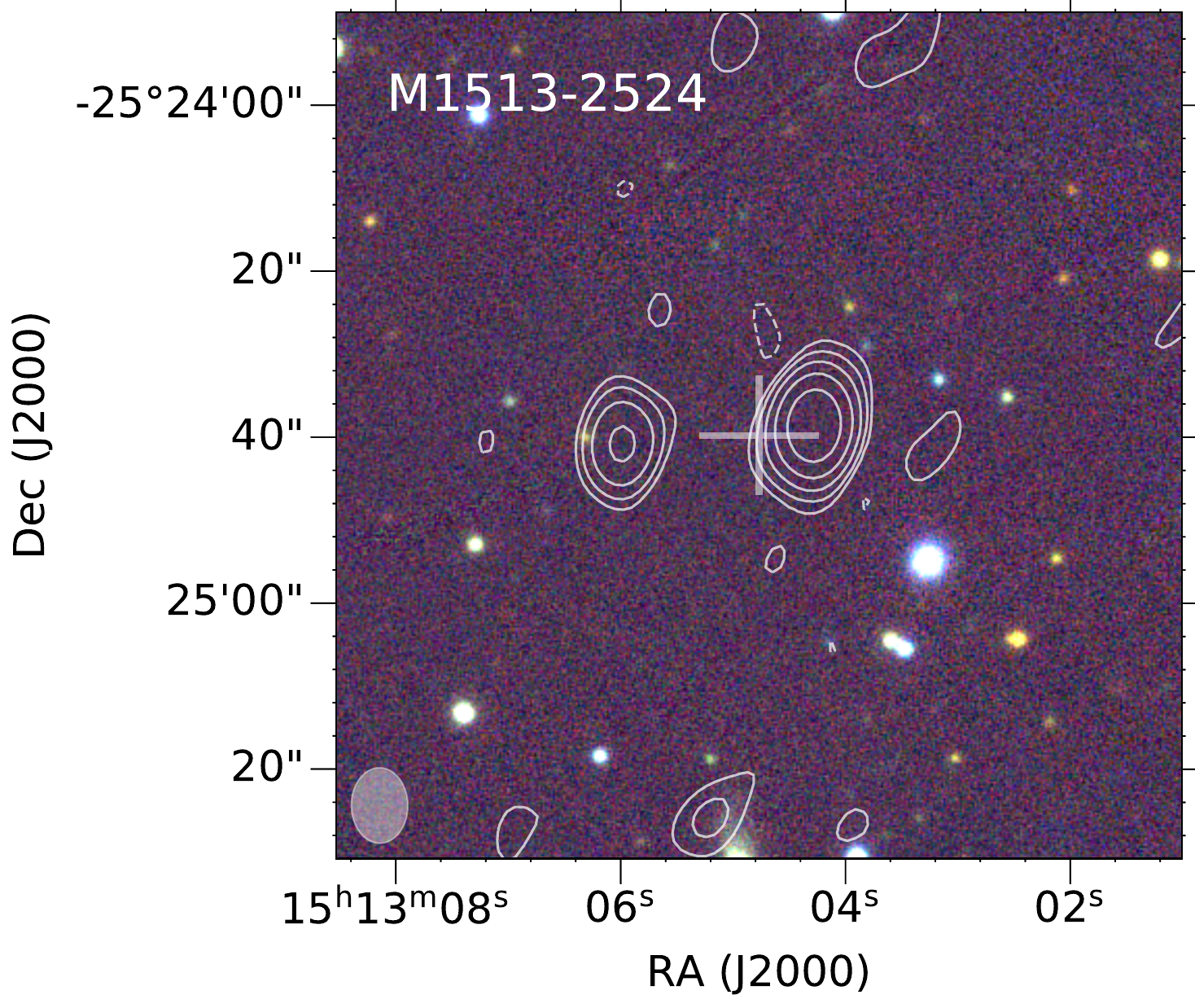}  
}}
}}  
\vskip+0.0cm  
\caption{
uGMRT radio continuum (420\,MHz) contours overlaid on PS1 $yig$ color composite images. For M0636-3106 and M0652-3230, the background image is PS1 $i$-band and uGMRT 420\,MHz, respectively.  The contour levels are shown at 20$\times$(-1, 1, 2, 4, 8, ...)\,mJy\,beam$^{-1}$. The synthesized beams, shown at the bottom-left corner of images, and the peak and total flux densities are provided in column 5 - 7 of Table~\ref{tab:wisesamp}, respectively. The position of WISE source is marked with a cross.
} 
\label{fig:maps}   
\end{figure*} 
%
\subsection{Data analysis}      
\label{sec:dataan}   

All the data were edited, calibrated and imaged using the Automated Radio Telescope Imaging Pipeline ({\tt ARTIP}) following the steps described in \citet[][]{Gupta21}. 
After flagging and calibration, the spectral line processing of wideband {\tt Band-3} data was  sped up by partitioning the 200\,MHz bandwidth into four 50 or 60\,MHz wide spectral windows with an overlap of 10\,MHz between the adjacent windows.  These spectral windows covered: 300-360\,MHz, 350-410\,MHz, 400-460\,MHz and 450-500\,MHz.  The calibrated visibilities for each spectral window were processed separately (independently) for RFI flagging, continuum imaging and self-calibration, and continuum subtraction.  The continuum subtracted visibilities were imaged to obtain RR and LL spectral line cubes. For this a `common' synthesized beam corresponding to the lowest frequency spectral window was used. 

The narrow band datasets from {\tt Band-2} survey and {\tt Band-3} follow-up observations were processed as a single 4.17 or 6.25\,MHz wide spectral window, respectively.  
%

\subsubsection{Continuum analysis}
\label{sec:contan}   
%

For {\tt Band-3}, the spectral window covering 390-450\,MHz, hereafter identified through the central reference frequency of 420\,MHz, is least affected by RFI resulting in best possible continuum images from the data.   
We used {\tt CASA} task {\tt IMFIT} to model the radio continuum emission in these images as multiple Gaussian components.  The 9  cases requiring more than one Gaussian component are shown in Fig.~\ref{fig:maps}.  
Only in 4 cases i.e., M0652$-$3230, M0910$-$0526, M1153$-$0442 and M1513$-$2524, does the second component contains more than 20\% of the total flux density.

In columns 5 and 6 of Table~\ref{tab:wisesamp} we list synthesized beams and peak flux densities of the most prominent Gaussian component.  For the above-mentioned four sources the second component is also listed. In the remaining cases the additional components are too faint ($<$50\,mJy) to be useful for the objectives of this paper, hence we do not list their individual properties. 
The total flux density as estimated from the single or multiple component fit is provided in column 7.  

\begin{figure} 
\centerline{\vbox{
\centerline{\hbox{ 
\includegraphics[trim = {0cm 0cm 0cm 0cm}, width=0.45\textwidth,angle=0]{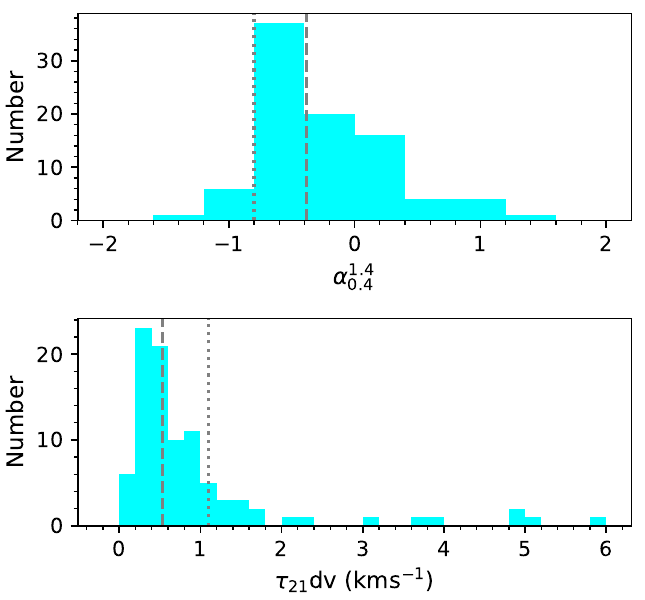}  
}} 
}}  
\vskip+0.0cm  
\caption{
Distributions of radio spectral index between 1400 and 420\,MHz ($\alpha^{1.4}_{0.4}$) and 5$\sigma$ 21-cm optical depth ($\int\tau$dv). 
The vertical dashed lines mark the median for each distribution.  
The dotted lines correspond to $\alpha$ = -0.8 and $\tau_{21}$dv = 1.1\,\kms.
} 
\label{fig:alphatdv}   
\end{figure}

In Fig.~\ref{fig:maps}, we also show optical images from Pan-STARRS1 (PS1) \citep[][]{Chambers16}.  Note that M0652-3230 is too south to be covered in PS1. The location of MIR sources from WISE are also shown in the images. Owing to the MIR-selection wedge described in Section~\ref{sec:samp}, all but one radio source in our sample are  quasars (see Section~\ref{sec:sampdef} for details). Indeed the median spectral index\footnote{Spectral index $\alpha$ is defined by the power law, $S_\nu \propto \nu^\alpha$}, $\alpha^{1.4}_{0.4}$, derived using the NVSS 1420\,MHz and the uGMRT 420\,MHz total flux densities  is $-0.38$ (see column 9 of Table~\ref{tab:wisesamp} and Fig.~\ref{fig:alphatdv}). As expected this is flatter than the overall radio source population which has $\alpha\sim-0.8$.
Thus, for our sample, when radio emission is dominated by a single component, we expect AGN to be located close to the peak of the radio emission.  In case two prominent radio components are present i.e., a compact symmetric object \citep[CSO;][]{Conway02} morphology, the AGN is expected to be located in between them. 
In all but 2 cases (M0910-0526 and M1513-2524; details provided below), the optical / MIR counterpart is at the location of the AGN expected from the radio morphology.   
As previously mentioned we have also verified these coincidences using higher spatial resolution 3\,GHz VLASS images.

In the case of M0910-0526, the northern component could be an unrelated radio source (Fig~\ref{fig:maps}).  We will exclude this component from the absorption line statistics.  M1513-2524, the only radio galaxy in the sample presented here, is among the optically faintest ($r>23$\,mag) in our survey.  
Among the two radio components, one of them is closer to the MIR source (see Fig~\ref{fig:maps}).  We have tentatively detected faint radio emission i.e., radio core at the location of the MIR source. For details see higher spatial resolution radio images presented in \citet[][]{Shukla21}.  We will consider the eastern and western radio components as the two radio lobes of this radio galaxy.   

For M1312$-$2026, the only target also observed in {\tt Band-2},  the properties at 234\,MHz are also provided in Table~\ref{tab:wisesamp}. The associated radio emission is compact with a deconvolved size of 15.8$^{\prime\prime}$$\times$13.8$^{\prime\prime}$ (position angle = $-38.0^{\circ}$).
Based on the observed flux densities, M1312-2026, has a spectral luminosity of L(1.4\,GHz) = $1.2\times10^{29}$\,W\,Hz$^{-1}$, which is more than three orders of magnitude higher than the radio power cut-off that separates FRI and FRII radio sources, 
and greater than the luminosity of any known radio-loud AGN at $z > 5$. The multi-frequency VLA and VLBA observations of this AGN have been obtained to investigate its radio morphology.

%
\subsection{Spectral line analysis}      
\label{sec:specan}   

\begin{figure*} 
\centerline{\vbox{
\centerline{\hbox{ 
\includegraphics[trim = {0cm 0.0cm 0cm 0cm}, width=1.0\textwidth,angle=0]{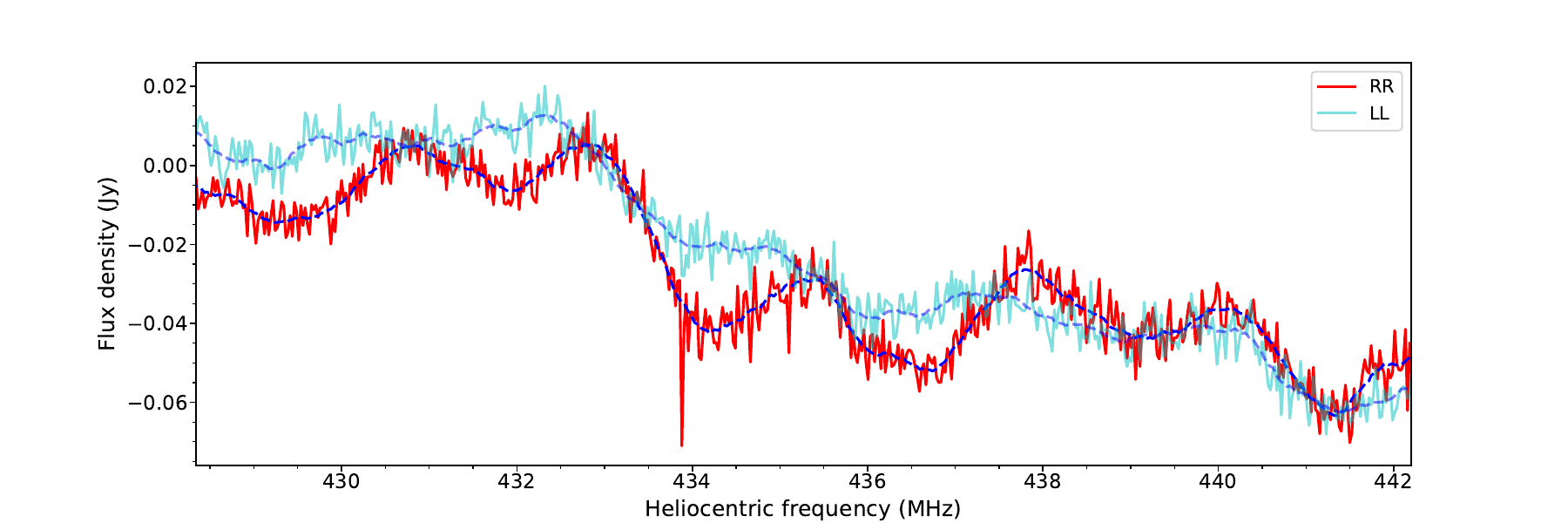}  
}} 
}}  
\caption{
	RR and LL spectra of M1206-0714 and continuum fits (dashed line).
} 
\label{fig:wiggle}   
\end{figure*} 
%

%
\begin{figure*} 
\centerline{\vbox{
\centerline{\hbox{ 
\includegraphics[trim = {1.0cm 3.0cm 1.0cm 2.0cm}, width=0.85\textwidth,angle=0]{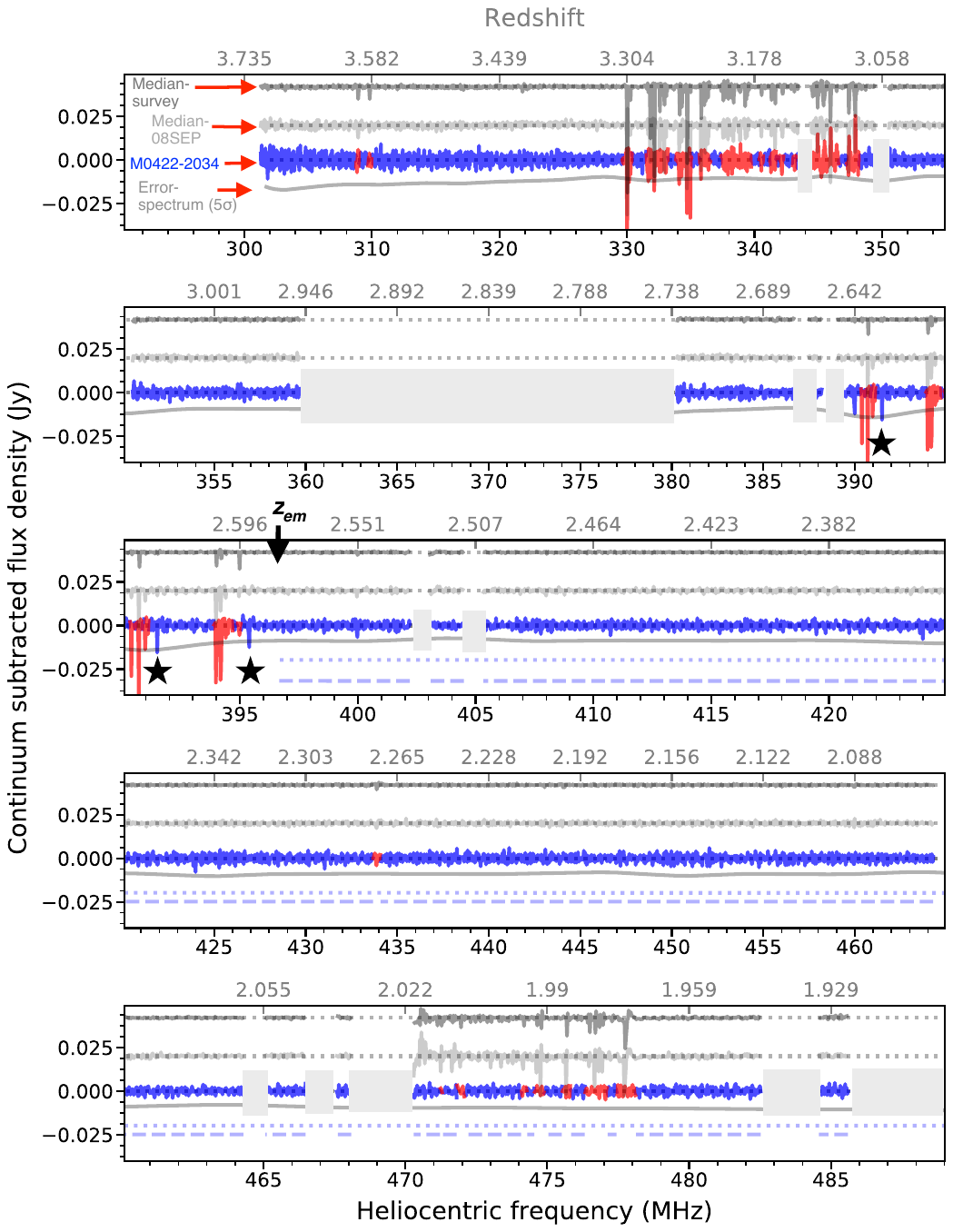}  
}} 
}}  
\vskip+0.0cm  
\caption{
	The continuum subtracted Stokes-$I$ spectrum of M0422-2034 ($z_{em}$=2.582; see arrow at 396.51\,MHz marking the redshifted \hi\ 21-cm line frequency).  
	Shaded regions mark frequency ranges that were masked prior to any calibration.
	The median spectra obtained using the full survey and only 08SEP run are plotted at +0.042 (median-survey) and +0.020\,Jy (median-08SEP), respectively. In the spectrum of M0422-2034, the pixels flagged on the basis of median spectra are shown in red.
	The  error spectrum (5$\times\sigma_{\rm rolling}$) is also shown.  
	The dotted and dashed lines at the bottom in panels 3-5 show the frequency range valid for 21-cm line search and actually contributing to the sensitivity function ($g(z)$), respectively. The candidate detections are marked using $\star$.
} 
\label{fig:fullspec}   
\end{figure*} 

For spectral line analysis, RR and LL spectra in the {\it heliocentric} frame were extracted at the location of radio continuum peaks from the spectral line cubes. The spectra show systematic oscillations or ripples due to residual bandpass calibration, and numerous positive/ negative spikes (for an example see Fig.~\ref{fig:wiggle}).  The ripple is not identical in the two parallel hands and also varies from one target source to other. 
We removed its effect by iteratively fitting the underlying structure using Savitsky-Golay filtering with {\tt window length = 24} and {\tt polynomial order = 3}.
In each iteration, the pixels deviating beyond the threshold were flagged and excluded from the subsequent iterations. The continuum was interpolated across the masked pixels and the process was repeated until no deviant pixels were found.

The above determined continuum fit was subtracted from the RR and LL spectra, and an error spectrum was generated by calculating a rolling standard deviation ($\sigma_{\rm rolling}$; window size=48 channels).  For {\tt Band-3}, an additional step was to merge the spectra from adjacent spectral windows and unmask the spikes to obtain the final RR and LL spectra covering the entire 300-500\,MHz.  These were then averaged with appropriate statistical weights to obtain the final Stokes-$I$ spectra. 
The resultant Stokes-$I$ spectra have flat baselines but numerous positive and negative spikes (for example see Fig.~\ref{fig:fullspec}).

\begin{figure} 
\includegraphics[trim = {0.0cm 12.0cm 0.0cm 10.0cm}, width=0.5\textwidth,angle=0]{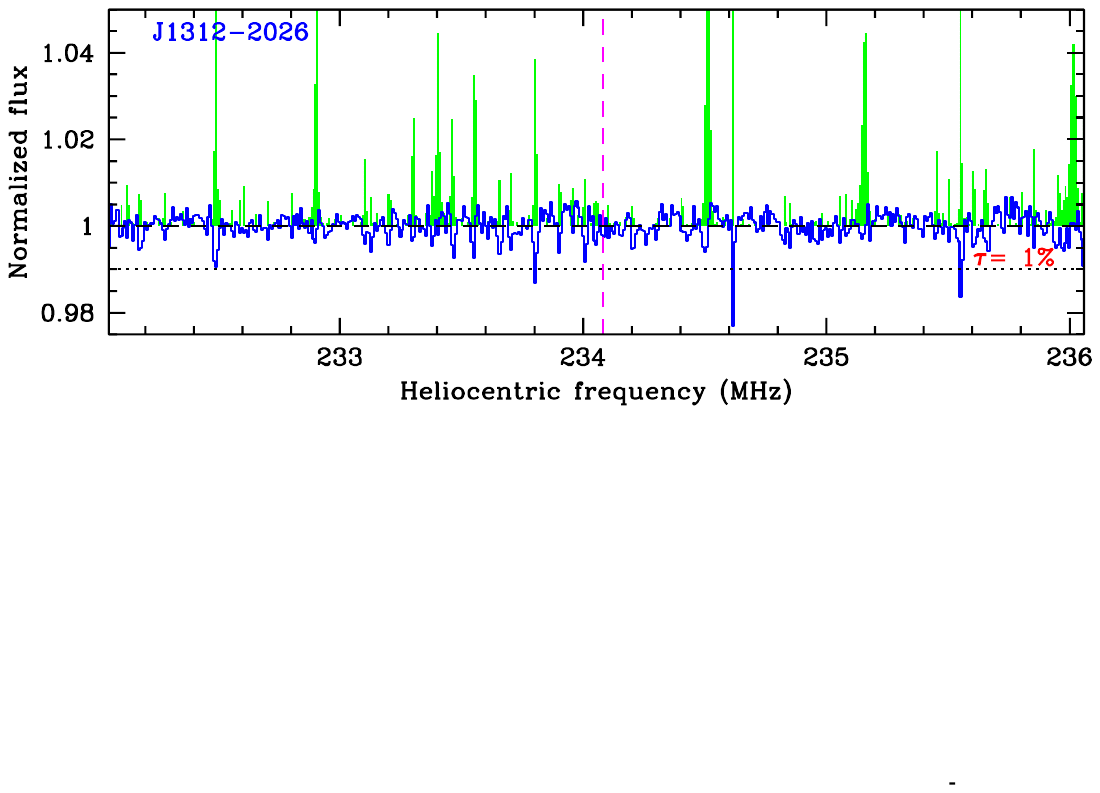}  
\caption{ \hi\ 21-cm absorption spectrum towards the highest redshift quasar M1312-2026 in our sample. The vertical dashed line marks the redshifted \hi\ 21-cm 
absorption frequency corresponding to \zq.  The filled histogram in the background is the ratio of the extent of data 
flagged due to frequency-dependent and frequency-independent flags.  
} 
\label{fig:j1312spec}   
\end{figure}

The {\tt Band-2} spectrum of M1312$-$2026 is presented in Fig.~\ref{fig:j1312spec}. These data were severely affected by the radio frequency interference (RFI).  The broad-band RFI mostly affected the shorter baselines ($<$4\,k$\lambda$) which were completely flagged.  There was also narrow-band impulse-like bursts of RFI, which affected all the baselines. Overall $\sim$55\% of the data was flagged due to antenna/baseline-based problems and RFI.  
The spectral rms in the Stokes-$I$ spectrum presented here  is 8.5\,mJy/beam, which for the unsmoothed spectrum presented here corresponds to a 1$\sigma$ optical depth sensitivity of 0.003. There are several statistically significant narrow-band features with widths of 1-2 spectral channels detected in the spectrum.  But all of these are coincident with the spikes present in the RFI-spectrum, and are most likely due to the low-level narrow-band RFI which could not be detected in the individual baselines.

In general, no true emission is expected in our spectra and only a tiny fraction of negative spikes are expected to represent true absorption. The majority of these spikes are RFI artefacts.  The biggest issue in spectral line search at low frequencies is to distinguish between true absorption and RFI artefacts.
The rest of this section is concerned with absorption line search in {\tt Band-3} spectra.

The worst RFI in {\tt Band-3} spectra (e.g., at 360-380\,MHz) was flagged by applying an initial mask prior to any calibration (see shaded regions in Fig.~\ref{fig:fullspec}).  
Further, to identify artefacts due to weaker but persistent RFI, we generated median Stokes-$I$ spectra for each observing run and the full survey. The median spectrum from the survey and 08SEP run in which M0422-2034 was observed are shown in Fig.~\ref{fig:fullspec}.  
The pixels deviating by more than 5$\sigma$ in the median spectra were taken to represent RFI artefacts. We rejected corresponding pixels in the individual source spectrum. In Fig.~\ref{fig:fullspec}, such pixels for M0422-2034 are plotted in red.  

After this we created a list of 550 absorption line candidates using a detection algorithm which requires: (i) flux density at a pixel $j$, F($j$) $< -5\times\sigma_{rolling}$($j$) and (ii) Heliocentric frequency at $j$, $\nu$($j$) $\ge \nu_{\rm 21cm}$/(1 + ($z_{em}$ + 0.01) ), where $\nu_{\rm 21cm}$ is the rest-frame 21-cm line frequency.  The factor of 0.01 in the denominator, which approximately corresponds to a outflowing velocity of $\sim$3000\,\kms, allows for the possibility of detecting redshifted  absorption associated with AGN \citep[see Fig.~21 of][]{Gupta06}.

\begin{figure}[b] 
\centerline{\hbox{
\includegraphics[trim = {0cm 1.0cm 0cm 0.0cm}, width=0.5\textwidth,angle=0]{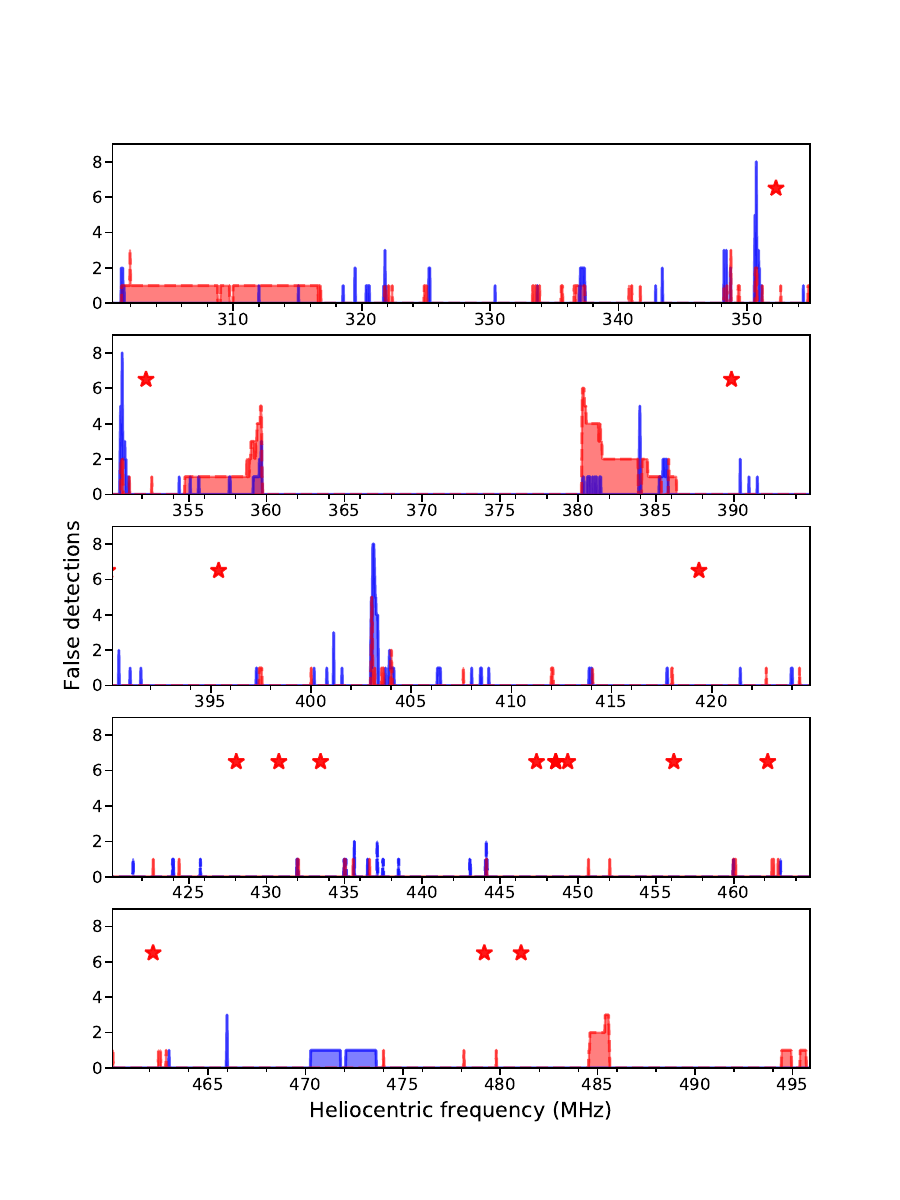} 
}}
\vskip+0.0cm  
	\caption{ Distribution of false absorption (blue) and emission (red) detections for the survey. The majority of these are the edges of frequency regions masked in Fig.~\ref{fig:m1540spec}. The locations of absorption candidates (see column 12 of Table~\ref{tab:wisesamp}) are marked by $\star$.
} 
\label{fig:falsedet}   
\end{figure}

Next, we created a false-detection spectrum by identifying pixels based on following two characteristics.  First, we identified all positive spikes with F($j$) $> 5\times\sigma_{rolling}$($j$).  These are unphysical and hence false detections because \hi\ emission lines are too weak to be detectable at $z>2$ in our spectra.  Second, we identified all negative spikes i.e., F($j$) $<-5\times\sigma_{rolling}$ but only at $\nu$($j$) $< \nu_{\rm 21cm}$/(1 + ($z_{em}$ + 0.01)).   These are unphysical because absorbing gas must be in the front of radio source. 
In Fig.~\ref{fig:fullspec}, we mark three candidates using $\star$. Two of these are clearly false detections whereas third one at 395.5\,MHz (approximately +800\kms\ with respect to $z_{em}$) could be a true absorption associated with the AGN. 
The cumulative distributions of all the false absorption (528) and emission (1359) detections from the survey are shown in Fig.~\ref{fig:falsedet}. 
These represent frequency ranges that may be affected by sporadic RFI. The majority of these are at the edges of frequency ranges masked in Fig.~\ref{fig:m1540spec}.
An updated RFI mask would get rid of them.  This will certainly be the preferred strategy for defining frequency ranges to be used for continuum imaging. 
Here, we rejected all the absorption candidates that are within one frequency channel of any of these false detections.  
This step reduced the number of absorption candidates by a factor of $\sim$10.

We visually examined RR and LL spectra of the remaining 48 absorption candidates for consistency. Specifically, we imposed the criteria that integrated optical depths estimated using the RR and LL spectra match within 3$\sigma$ and within errors the absorption profiles appear similar. 

\begin{deluxetable}{lcccc}
	\tablecaption{High probability 21-cm absorption candidates}
\tabletypesize{\scriptsize}
\tablehead{
	\colhead{Source name} &  \colhead{$z_{em}$} & \colhead{ $z_{abs}$(21-cm) } & \colhead{ $\int\tau$dv({21-cm}) }  & \colhead{ $\Delta$V$_{90}$ } \\ 
 			      &                     &     & \colhead{(\kms)} & \colhead{(\kms)} \\ 
	\colhead{  (1)   }    & \colhead{  (2)   } &  \colhead{  (3)   }            &     \colhead{  (4)   }    &     \colhead{  (5)   }           
} 
\startdata
	M0513$+$0100   & 2.673     &  1.9526     &  3.59$\pm$0.67 &   107              \\  
	M0618$-$3158   & 2.134     &  1.9642     &  0.49$\pm$0.07 &   15               \\ 
	M1244$-$0446   & 3.114     &  2.3871     &  1.61$\pm$0.27 &   70               \\ 
	M1312$-$2026   & 5.064     &  3.0324     &  0.34$\pm$0.08 &   20               \\  
	M1540$-$1453   & 2.098     &  2.1139     &  9.14$\pm$0.28 &   144              \\ 
\label{tab:abscand}
\enddata
\tablecomments{ 
Column 1: source name. Column 2: emission redshift. Columns 3-5: \hi\ 21-cm absorption redshift, integrated 21-cm optical depth limit and velocity width of absorption profile based on the survey spectra presented in Fig.~\ref{fig:21cmspec}, respectively. The spectra have been normalized using the corresponding peak flux densities.
}
\end{deluxetable}
%
After all the statistical filtering described above, we were left with a total of 15 candidates towards the sight lines which are identified in column 12 of Table~\ref{tab:wisesamp}.  We extracted Stokes-$I$ spectra of gain calibrators corresponding to these. For the following candidates:  M0212-3822 (\zabs=2.1666), M0250-2627 (\zabs=2.1665), M0422-2034 (\zabs=2.5924), M0513+0100 (\zabs=2.1612, 2.3183, 2.6434), M0515-0120 (\zabs=2.1753), M0702-3302 (\zabs=2.2769), M1448-1122 (\zabs=2.2973) and M2017-2933 (\zabs=2.0733), we find an `absorption' feature at the same redshifted frequency in the gain calibrator spectrum. The angular separation between a target source and its gain calibrator is typically 15\,degree.    
Thus, it is unrealistic that at $z>2$ a true absorption is present in both of them.  Therefore we rejected these 10 candidates.   

\begin{figure} 
\centerline{
\vbox{
\centerline{\hbox{ 
\includegraphics[trim = {0cm 7cm 0cm 0cm}, width=0.25\textwidth,angle=0]{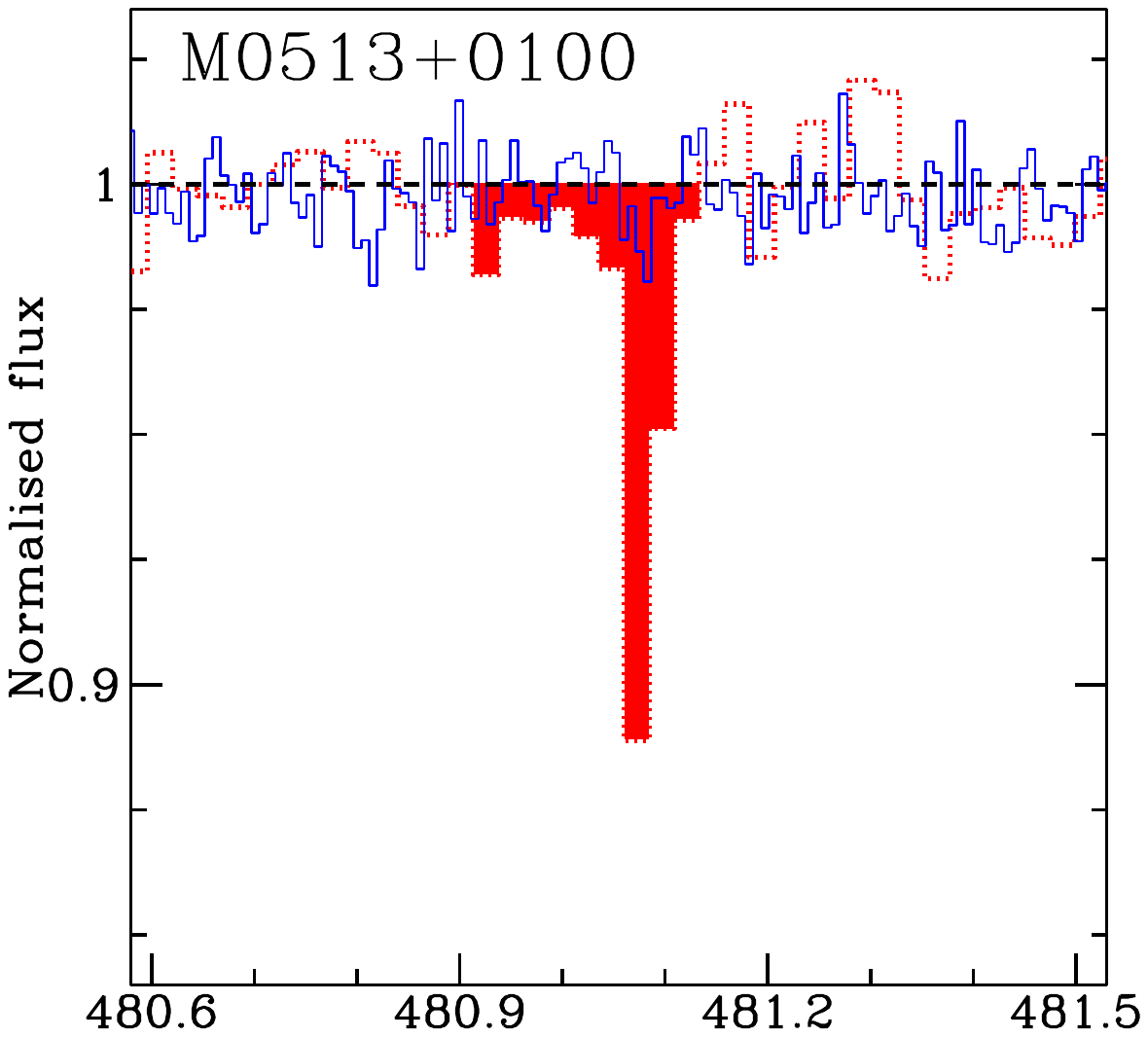}  
\includegraphics[trim = {0cm 7cm 0cm 0cm}, width=0.25\textwidth,angle=0]{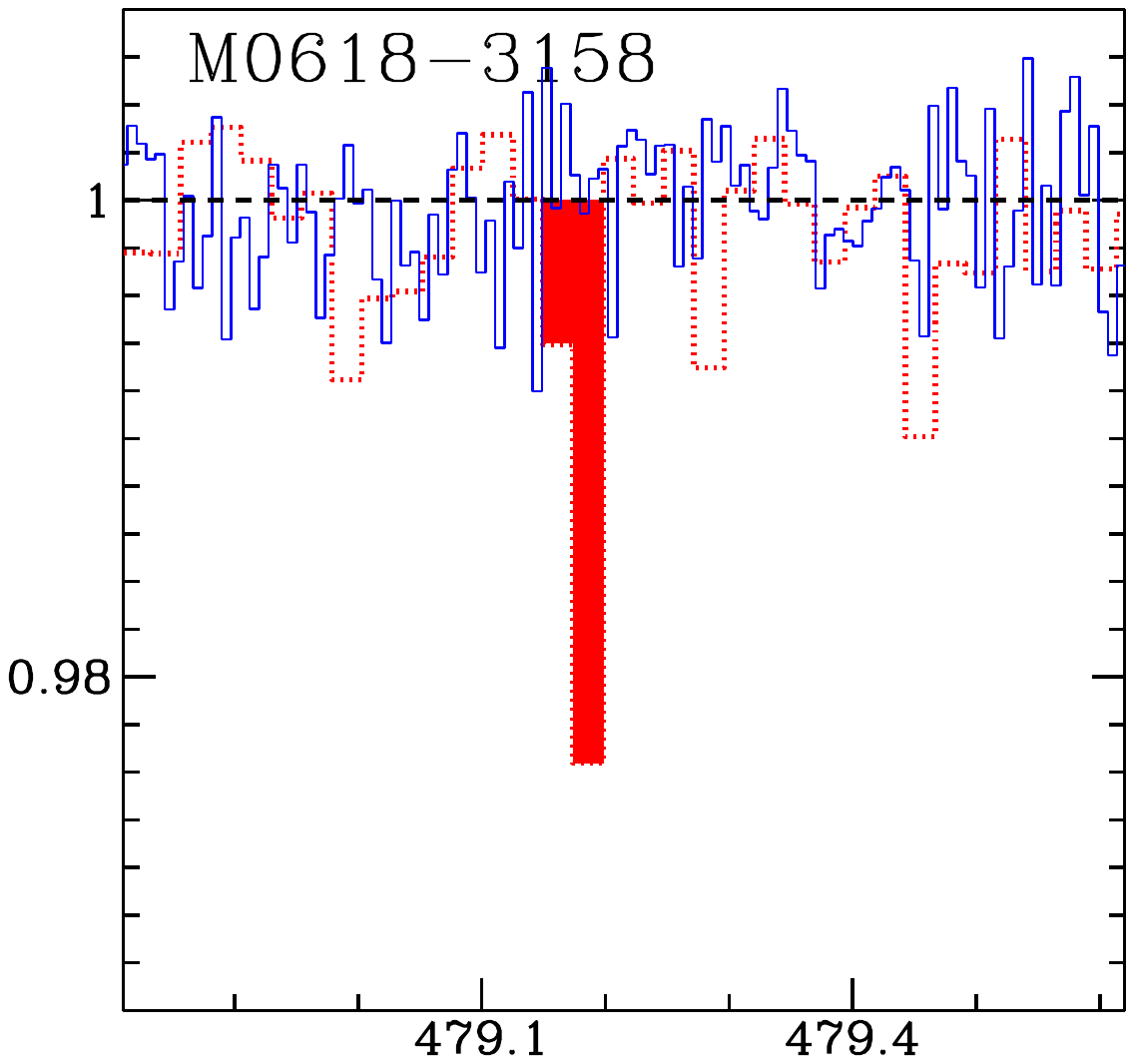} 
}}
\centerline{\hbox{
\includegraphics[trim = {0cm 6cm 0cm 0cm}, width=0.25\textwidth,angle=0]{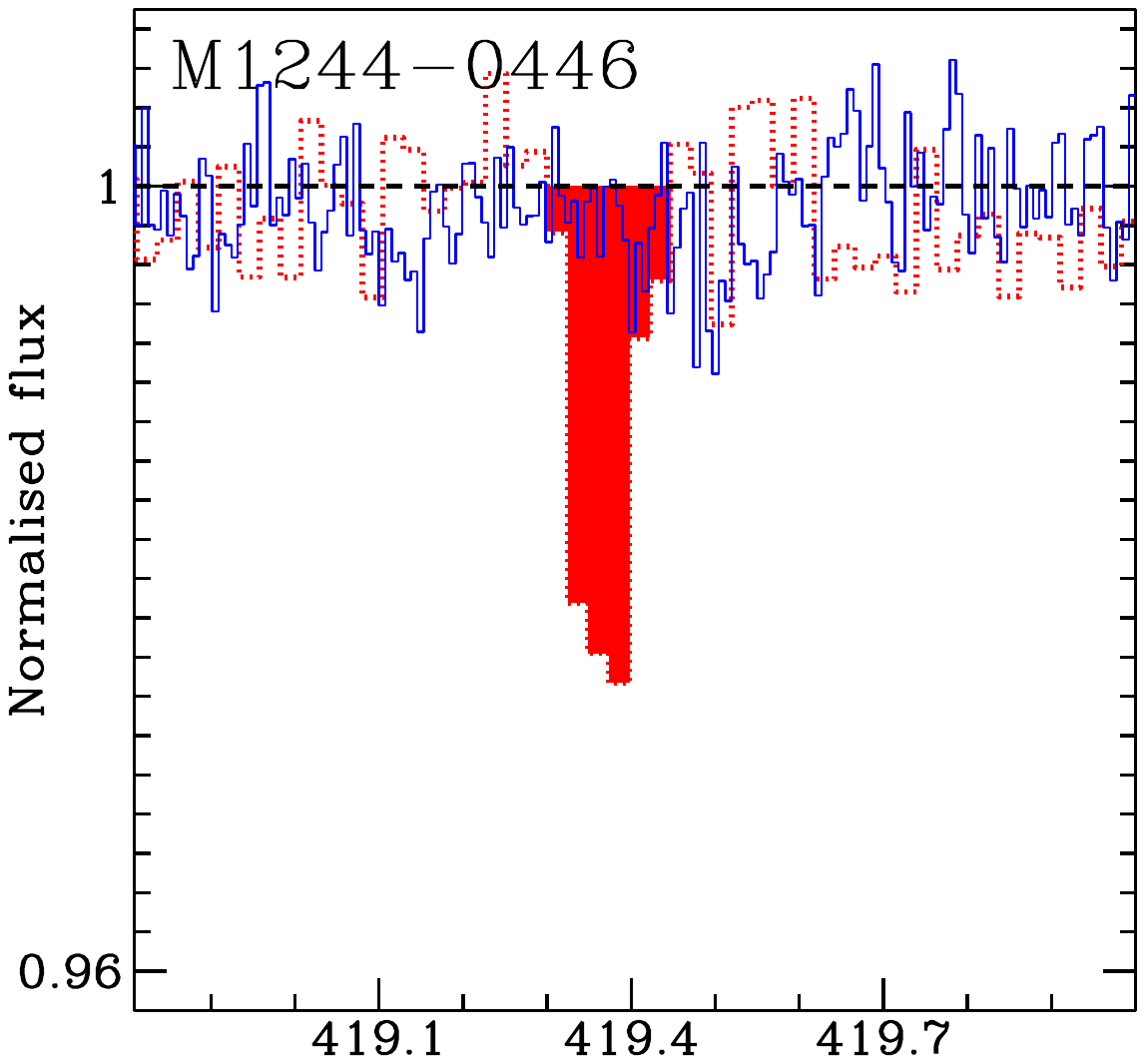}  
\includegraphics[trim = {0cm 6cm 0cm 0cm}, width=0.25\textwidth,angle=0]{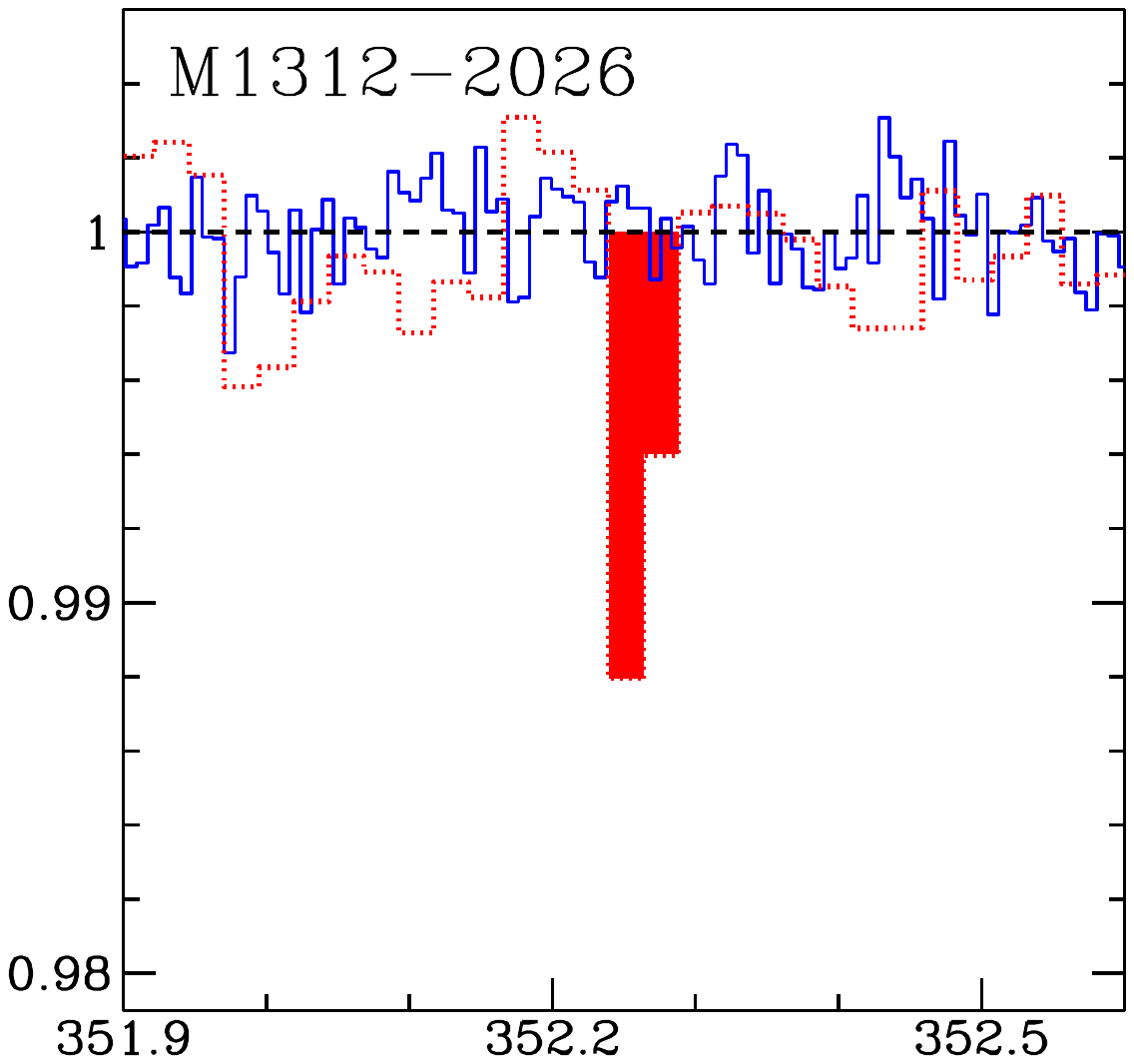}  
}}
\centerline{
\includegraphics[trim= {0cm 6cm 0cm 0cm}, width=0.25\textwidth,angle=0]{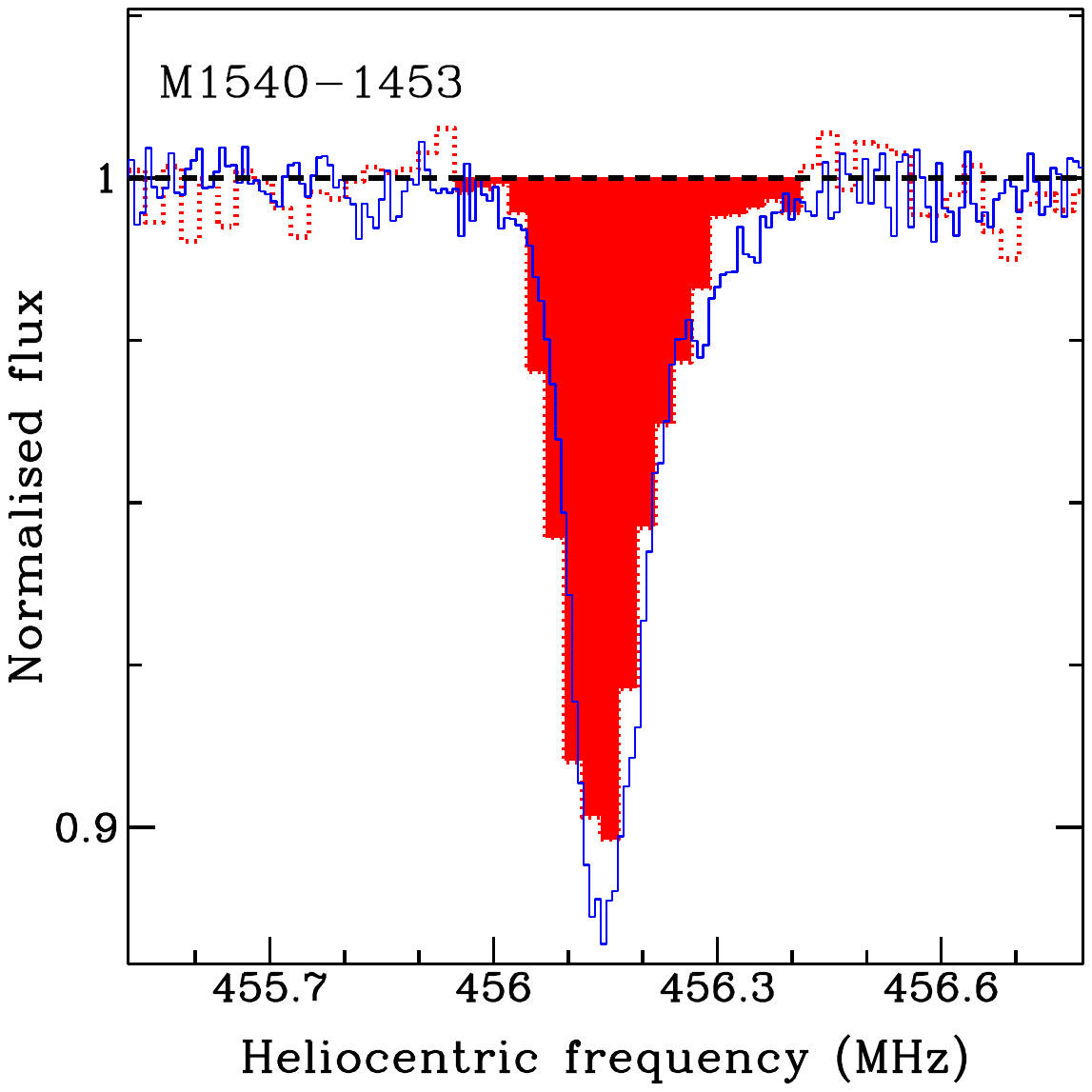} 
}
}
}  
\caption{High probability absorption candidates.  The survey and reobservation spectra are shown as dotted (red) and solid (blue) lines, respectively. 
} 
\label{fig:21cmspec}   
\end{figure} 

Finally, we have 5 high probability candidates. These are listed in Table~\ref{tab:abscand}.  We also estimated integrated optical depths ($\int\tau$dv) and velocity width ($\Delta$V$_{90}$) corresponding to the 5\% and 95\% percentiles of the apparent optical depth distribution.  These are very similar to the values observed for 21-cm absorption lines detected in various surveys \citep[see e.g.,][]{Gupta09, Dutta17}.

We reobserved these high probability candidates with uGMRT (see Section~\ref{sec:obs} and Table~\ref{tab:obslog}) using a bandwidth of 6.25\,MHz centered at \hi\ 21-cm line frequency corresponding to $z_{abs}$(21-cm) given in column 3 of Table~\ref{tab:abscand}.  These observations were carried out at night to reduce the effect of RFI.  
For better RFI mitigation, the frequency setup was chosen to provide a spectral resolution of $\sim$1\,\kms. Recall, the survey observations had a spectral resolution of $\sim$18\,\kms.  
In Fig.~\ref{fig:21cmspec}, we present profiles from the survey and reobservation spectra.  Clearly, only M1540-1453 is confirmed. The remaining 4 candidates are due to RFI.

To summarize, based purely on the uGMRT survey spectra and blind 21-cm line search, we identified 5 absorption features (4 intervening and 1 associated systems).  The followup observations confirmed only one of these i.e., absorption associated with the radio source M1540-1453 at $z_{em}$ = 2.098. 
The distribution of 5$\sigma$ 21-cm optical depth limits  at 420\,MHz estimated assuming $\Delta v$=25\,\kms\ is shown in Fig.~\ref{fig:alphatdv}.  The median 0.535\,\kms\ is well below the sensitivity (1.1\,\kms) required to detect CNM (i.e T$\sim$100 K) in DLAs (i.e $N$(H~{\sc i})$\ge 10^{20.3}$\,\cmsq).

\section{Associated \hi\ 21-cm absorption detection towards M1540-1453}    
\label{sec:dets}  

\begin{figure} 
\centerline{
\includegraphics[trim= {0cm 5cm 0cm 4cm}, width=0.5\textwidth,angle=0]{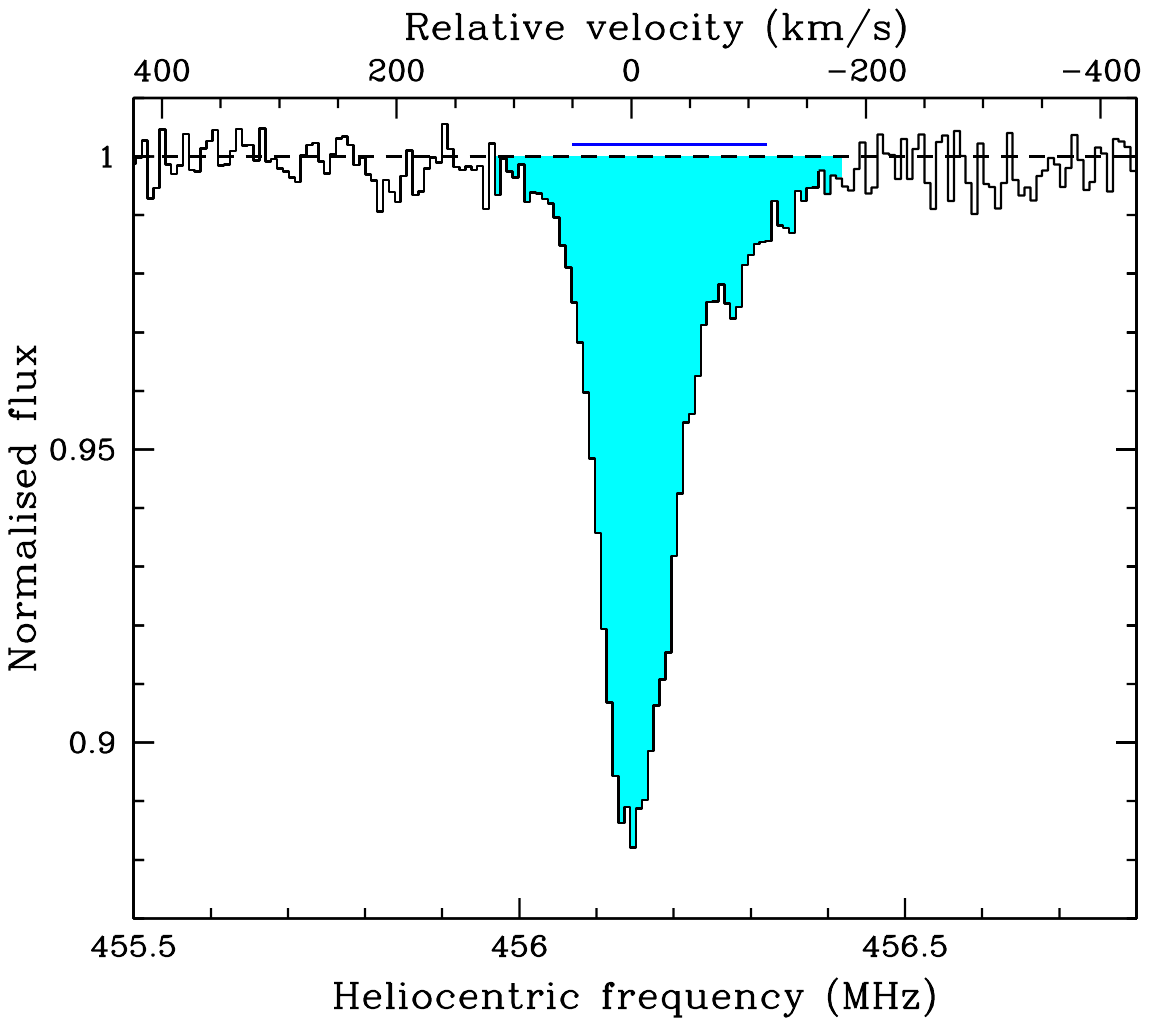}  
}  
\caption{Associated \hi\ 21-cm absorption detection towards M1540-1453.  
The zero of the velocity scale is defined with respect to the peak of the absorption i.e., \zabs = 2.1139. The solid horizontal line corresponds to $\Delta V_{90}$.
} 
\label{fig:m1540spec}   
\end{figure} 

The \hi\ 21-cm absorption spectrum of M1540-1453 based on the followup observation is presented in Fig.~\ref{fig:m1540spec}.  
It is normalized using the corresponding peak continuum flux density of 652\,mJy\,beam$^{-1}$.
The absorption is spread over $\sim$ 300 \kms. We measure a total optical depth, $\int \tau $dv = 11.30$\pm$0.07\,\kms\ and about 90\% of this is confined to $\Delta V_{90}$ = 167\,\kms. This translates to a \hi\ column density of $(2.06\pm0.01)\times 10^{21} ({T_S\over 100}) ({1\over f_c})$ cm$^{-2}$.  Here, $T_{\rm s}$ is the spin temperature in Kelvin and $f_c$ is covering factor of absorbing gas.  
We note that measuring the covering factor of the absorbing gas would require milliarcsecond scale spectroscopy which is currently not feasible at $z >0.2$ \citep[][]{Srianand13dib, Gupta18j1243}. Therefore, unless explicitly stated otherwise, hereafter, we will assume $f_c= 1$. We also assume $T_{\rm s}$ = 100\,K for the CNM \citep[][]{Heiles03}.

We obtained an optical spectrum of M1540-1453 using the Robert Stobie Spectrograph \citep[RSS; ][]{Burgh03, Kobulnicky03} on SALT as part of our optical survey summarized in Section~\ref{sec:samp}.  The spectrum has a typical signal-to-noise ratio (SNR) of 7 per pixel. It shows C~{\sc iv} and [C~{\sc iii}] emission lines. The spectral SNR close to the [C~{\sc iii}] emission line is poor due to residuals from the subtraction of skylines. Hence, we focus on the C~{\sc iv} emission line. The peak of \civ\ emission corresponds to \zem $\simeq$ 2.113 which is consistent with the 21-cm absorption peak (Fig.~\ref{fig:m1540spec}). The emission line is superimposed with absorption lines possibly also of \civ\ but at a redshift slightly lower than the 21-cm absorption.
%

Our SALT spectrum does not cover \lya\ absorption for M1540-1453.  But it covers the rest wavelength range of 1436 to 2414\AA. Although the region is affected by skylines, in principle, we have access to the Fe~{\sc ii} lines associated with the 21-cm absorption. 
We detect an absorption feature exactly at the redshifted wavelength corresponding to Fe~{\sc ii}$\lambda$2383 line. The redshift and rest equivalent width of the absorption are \zabs = 2.11404 and $W_{\rm r}$ = $1.05\pm0.25$\AA, respectively.  We also find absorption dips at the expected positions of the Fe~{\sc ii}$\lambda$2344, Fe~{\sc ii}$\lambda$2374 and Si~{\sc ii}$\lambda$1526 lines. These coincidences are interesting because metal absorption line ratios can be a reasonable indicator of \hi\ column density \citep[][]{Rao06, Gupta12, Dutta17fe2}.  
We note that the rest frame ultraviolet luminosity of the quasar at 912\AA\ i.e., $L_{912}$ = $1.4\times10^{23}$\,W\,Hz$^{-1}$ (see Section~\ref{sec:assoc} for details). This is marginally above the ultraviolet cut-off of $10^{23}$\,W\,Hz$^{-1}$ above which \hi\ 21-cm absorption is rarely detected \citep[][]{Curran10assoc}. 
A better quality optical spectrum covering Ly$\alpha$ and above-mentioned metal lines is needed to extract physical conditions prevailing in the absorbing gas.  

We follow the method described in \citet{Srianand08bump} to constrain the visual extinction, $A_V$.  Using our flux calibrated SALT spectrum along with the Small Magellanic Cloud (SMC) type extinction curve and the average QSO spectral energy distribution (SED) given in \citet{Selsing16}, we measure, $A_V = 0.13\pm0.01$. 
The moderate extinction observed towards M1540-1453 is consistent with the idea that cold atomic gas is accompanied by dust \citep[see Fig.~9 of][]{Dutta17fe2}.

To date only three associated \hi\ 21-cm absorbers are known\footnote{We note that \hi\ 21-cm absorber ($z_{abs}$ = 1.9436) has been detected towards the QSO PKS\,1157+014 at $z_{em} = $ 1.978 \citep[][]{Wolfe81}  which is slightly below the redshift cut-off used here.  It is suggested that in this case the absorption originates from a galaxy bound to the cluster containing the QSO.} at $z>2$. These are: \zabs = 3.3968 towards B2~0902+345 \citep{Uson91,Briggs93},  \zabs = 2.6365 towards MG\,J0414+0534 \citep{Moore99} and \zabs = 3.52965 towards 8C\,0604+728 \citep{Aditya21}. Thus, M1540-1453 absorber reported here is only the fourth detection at $z>2$.
The inferred column densities for reasonably assumed values of spin temperature and covering factor ($T_{\rm s}$ = 100\,K; $f_c$ =1) imply $N$(\hi)$>>2\times 10^{20}$\,\cmsq\ which is the formal DLA cut-off.  So, in all these cases if the optical and radio sightlines coincide then one expects to see a DLA at the 21-cm absorption redshift. 
We investigate this for B2~0902+345 and MG~J0414+0534, the two sources for which optical spectra are available in literature.

B2~0902+345 is associated with a radio galaxy that exhibits  \lya\ emission extended up to 50~kpc.  The associated radio continuum emission ($\alpha$ = -0.94) has a highly distorted radio morphology over 6$^{\prime\prime}$ ($\sim$45\,kpc at $z_{abs}$) and rotation measure in excess of 1000\,rad\,m$^{-2}$ \citep[][]{Carilli94}. However, no signatures of \lya\ absorption associated with the 21-cm absorber are seen. In fact, the 21-cm absorption is found to be redshifted with respect to the \lya\ emission  \citep[shift $\sim$ +300\,\kms; see][for details]{Adams09}. 
%

MG~J0414+0534 is a highly reddened gravitationally lensed quasar \citep[$A_V\sim$5][]{Lawrence95}. The weakness of \lya\ emission prevents us from searching for a DLA. However 4 associated strong Fe~{\sc ii} absorption components are detected in the redshift range 2.6317-2.6447 \citep[][]{Lawrence95abs}. These are within the range over which CO emission is detected but do not exactly coincide with the 21-cm absorption redshift, which itself is shifted by $\sim$200\,\kms\ with respect to the peak of the CO emission line. The 21-cm absorption in this case may actually be towards a steep-spectrum radio jet component not spatially coinciding with the AGN \citep[][]{Moore99}.  The same scenario may also apply to B2~0902+345. 

In comparison, the radio emission associated with M1540-1453 is compact in VLASS and the radio continuum peak coincides well with the PS1/MIR counterparts. The deconvolved radio source size is $1.8^{\prime\prime}\times 0.7^{\prime\prime}$ with a position angle of $155^\circ$. This corresponds to an upper limit on the size of $\sim$10\,kpc at $z_{abs}$ = 2.1139. Clearly,  milliarcsecond scale imaging is required to estimate $f_c$ and understand the coincidence between 21-cm absorption and metal absorption lines observed towards this source.  Interestingly, the \hi\ 21-cm absorption is only slightly asymmetric at the base and does not show signatures of outflows i.e., blue-shifted absorption \citep[][]{Vermeulen03, Gupta06}.  In low-$z$ samples, such  absorption likely originates from the circumnuclear disk or gas clouds associated with the host galaxy \citep[e.g.,][]{Gereb15, Srianand15}. 
In Section~\ref{sec:assoc}, we note that the quasars in our sample are generally hosted in gas and dust poor galaxies.  The CO emission line and millimetre continuum observations of M1540-1453 will reveal the nature of its host galaxy ISM and shed light on the origin of gas detected in 21-cm absorption.

\section{Intervening absorption statistics}      
\label{sec:int}   

In this section, we constrain the occurrence of intervening \hi\ 21-cm absorbers at $z>2$ using a blind spectroscopic search.  We also use \lya\ and metal absorption lines detected in our SALT spectra to interpret these statistics.

\subsection{Blind \hi\ 21-cm absorption line search}      
\label{sec:intblind}   

To estimate the incidence of intervening absorbers, we first determine the sensitivity function, $g({\cal{T}}, z)$, as a function of integrated optical depth (${\cal{T}}$) and redshift ($z$).  For this we follow the formalism provided in \citet[][]{Gupta21} which takes into account the varying optical depth sensitivity across the spectrum. The two crucial inputs required to determine $g({\cal{T}}, z)$ are spectral weight ($W$) and completeness fraction ($C$).  The former accounts for the possibility that some of the targets in the sample may not have spectroscopic redshifts.  Since, all the targets in our sample have spectroscopic redshifts, we assign $W$ = 1.   

\begin{figure} 
\centerline{\vbox{
\centerline{\hbox{ 
\includegraphics[trim = {0cm 0cm 0cm 0cm}, width=0.5\textwidth,angle=0]{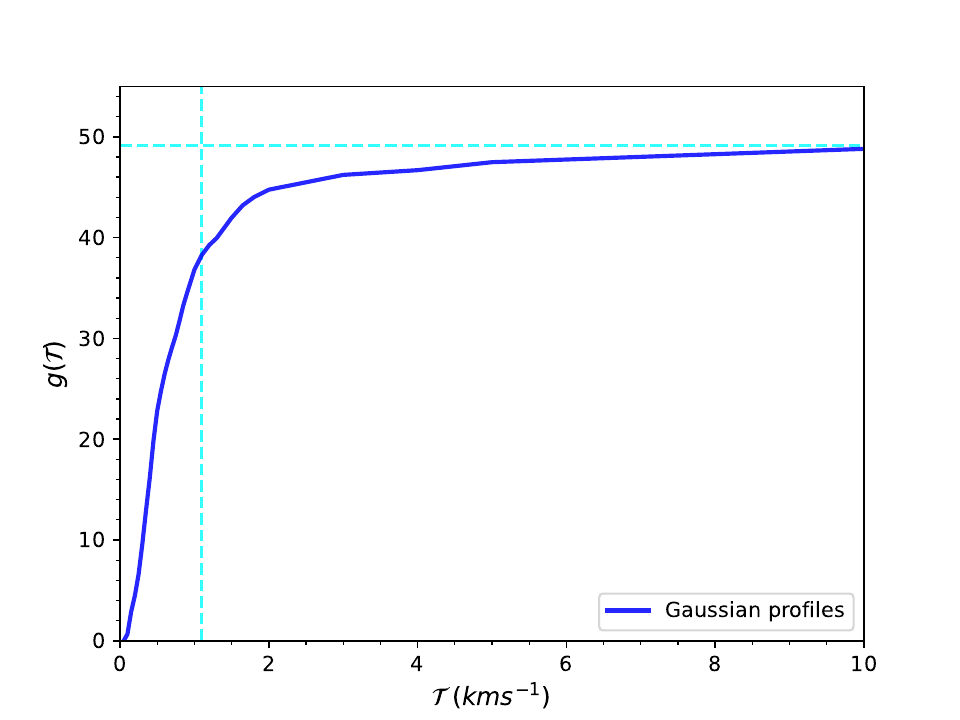}  
}} 
}}  
\vskip+0.0cm  
\caption{
	Completeness corrected total redshift paths ($\Delta z ({\cal{T}}) \equiv g({\cal{T}})$) for the 21-cm line search. 
	The horizontal dashed line represents total redshift path without completeness correction.
	The vertical dashed lines correspond to integrated optical depth, ${\cal{T}}$ = 1.1\,\kms.  This corresponds to a  5$\sigma$ detection limit of $N$(\hi)\,=\,$2\times 10^{20}$\,\cmsq\ for $T_s$ = 100\,K.  
} 
\label{fig:gzcompl}   
\end{figure} 

\begin{figure*} 
\centerline{\vbox{
\centerline{\hbox{ 
\includegraphics[trim = {0cm 0cm 0cm 0cm}, width=1.0\textwidth,angle=0]{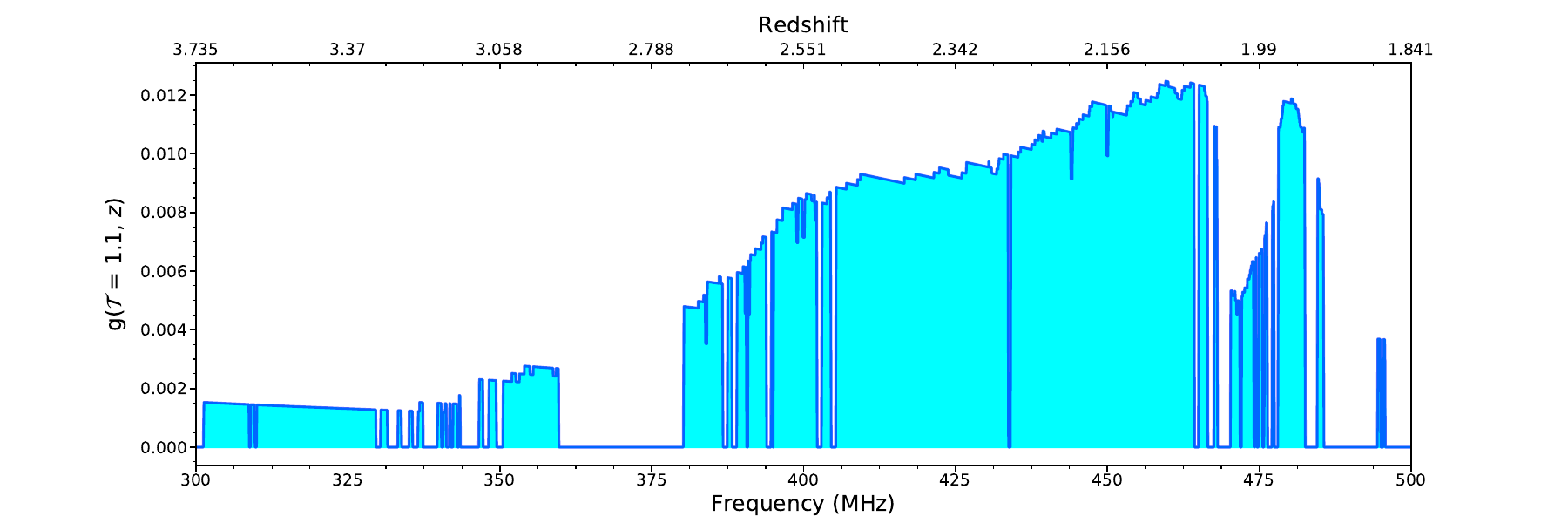}  
}} 
}}  
\vskip+0.0cm  
\caption{
	The sensitivity function, $g(z)$, for \hi\ 21-cm absorbers with integrated 21-cm optical depth ${\cal{T}} \ge$  1.1\,\kms.  The abrupt dips are caused by the spectral channels removed from the data due to RFI.  Approximately 30\% of the redshift path is lost due to RFI.
} 
\label{fig:gz}   
\end{figure*} 

The completeness fraction accounts for the detectability of absorption line features of different line shapes.  To determine this we consider the absorption profiles of all the intervening absorbers detected from our surveys in last 15 years. 
We inject 200 single Gaussian components with widths consistent with the distribution of $\Delta V_{90}$ of these absorbers \citep[see Fig.~8 of][]{Gupta21} at each pixel and apply the detection algorithm described in Section~\ref{sec:specan} to compute $C({\cal{T}}_j, z_k)$ as 
\begin{equation}
	C({\cal{T}}_j, z_k) = \frac{1}{N_{inj}}  \sum_{i=1}^{N_{inj}}  F({\cal{T}}_j, z_k, \Delta V_{i}), 
\label{eqgz2}
\end{equation}
where $N_{inj}$ is the number of injected systems and $F=1$ if the injected system is detected and 0 if not.
The total completeness corrected redshift path of the survey, $g({\cal{T}}_j$), considering all sight lines is plotted in Fig.~\ref{fig:gzcompl}. The redshift path starts falling off rapidly below ${\cal{T}}$ = 1.5\,\kms.  

It is of particular interest to consider the detectability of \hi\ 21-cm absorption in DLAs, i.e., ${\cal{T}}$ = 1.1\,\kms (refer to vertical dashed line in Fig.~\ref{fig:gzcompl}).  The sensitivity function providing the number of spectra in which it is possible to detect CNM in DLAs is shown in Fig.~\ref{fig:gz}.
The total redshift and comoving path length are $\Delta z = $ 38.3 and  $\Delta X =$ 130.1, respectively. 
Then, the incidence or number of 21-cm absorbers per unit redshift and comoving path length are $n_{21} <$  0.048 and $\ell_{21} < $  0.014, respectively. 
These 1$\sigma$ upper limits are based on small number Poisson statistics \citep[][]{Gehrels86}.
%

The formalism to search for \hi\ 21-cm absorption line presented above can also be applied to OH main lines. 
For stronger OH main line at 1667.359\,MHz, ${\cal{T}}$ = 1.1\,\kms\ and excitation temperature of 3.5\,K will correspond to  $N$(OH) = 8.6$\times10^{14}$\,\cmsq.  The total redshift and comoving path length are $\Delta z = $ 44.9 and  $\Delta X = $ 167.7, respectively. The number of OH absorbers per unit redshift and comoving path length are $n_{\rm OH}$ = 0.041 and $\ell_{\rm OH} < $  0.011, respectively.

Besides \hi\ 21-cm absorption, CNM at high-$z$ may also be searched using absorption lines of H$_2$ and C~{\sc i} \citep[e.g.,][]{Srianand12dla,Noterdaeme18}. Recently, \citet[][]{Krogager20} using a canonical two phase model of atomic gas and the observed statistics of H$_2$ and C~{\sc i} absorbers at high-$z$ estimated the comoving path length of CNM, $n_{\rm CNM}$ = 0.012.  The upper limit of $n_{21} < $  0.048 obtained through our blind survey is consistent with this. This result is the first study comparing the CNM cross-section of galaxies at $z>2$ estimated using radio and optical/ultraviolet absorption lines.  The detectability of H$_2$ and C~{\sc i} absorption at optical/ultraviolet and \hi\ 21-cm absorption at radio wavelength are affected by different systematic effects.  Indeed, there does not exist a one-to-one correspondence between the presence of H$_2$ and \hi\ 21-cm absorption, and the difference may be due to small sizes of H$_2$ bearing clouds \citep[see][for a discussion]{Srianand12dla}. Much larger radio surveys are needed to disentangle these possibilities.

\subsection{Relationship with \lya\ and metal lines}      
\label{sec:intopt}   

\begin{figure*}
\begin{center}
  \includegraphics[viewport=40 30 590 850,height=0.9\textheight,angle=270,clip=true]{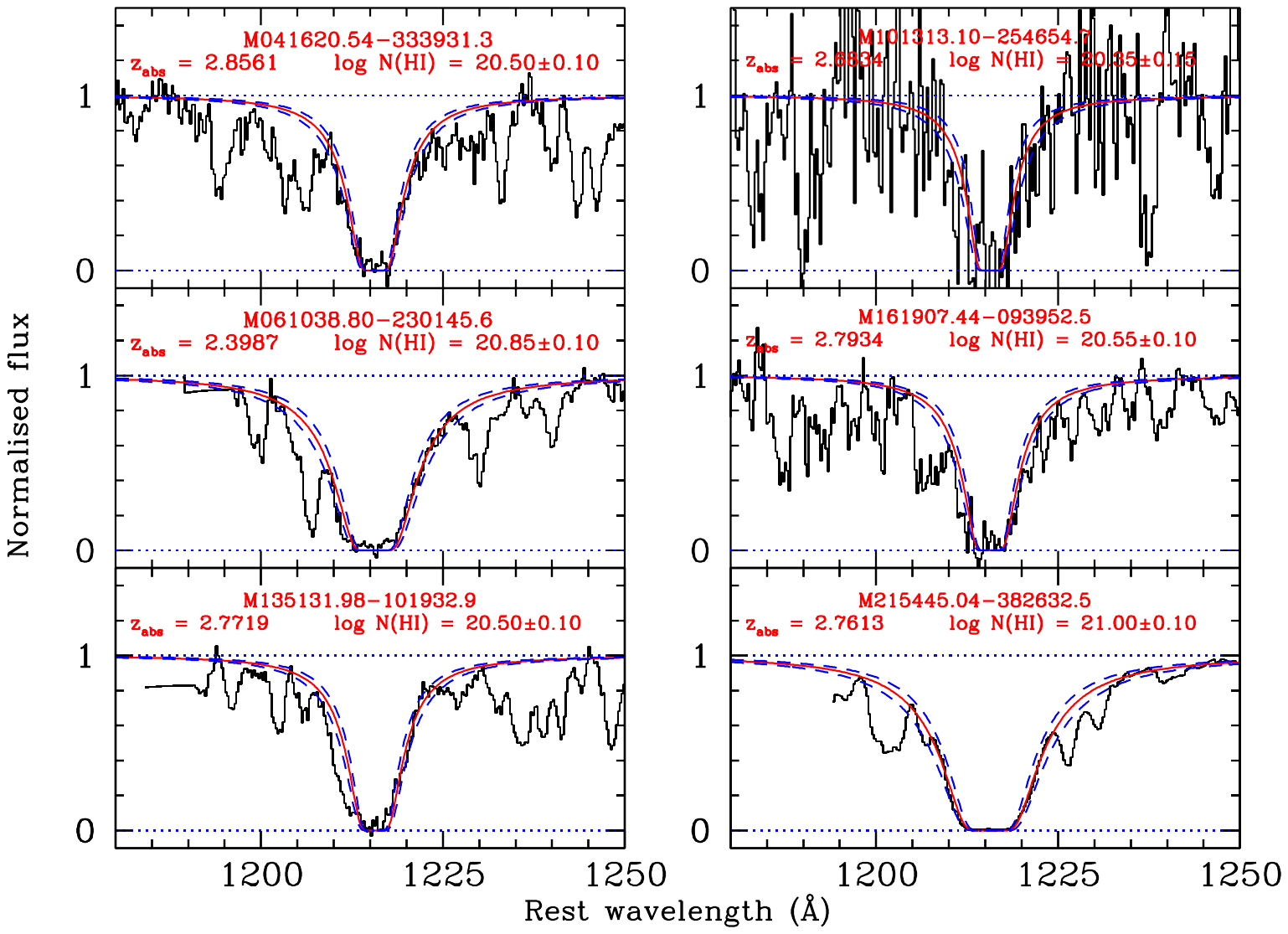}
\end{center}
\caption{Voigt profile fits to the 6 DLAs detected in our new radio loud quasar sample. The profiles
given in solid and long-dashed lines correspond to the best fit and 1$\sigma$ range around it respectively. The absorption redshift and the best fit $N$(H~{\sc i}) 
obtained are also provided in each panel. }
\label{fig_dlafit}
\end{figure*}

\begin{deluxetable*}{cccccccc}
\tablecaption{Properties of the DLAs derived from our observations}
\tablehead{
\colhead{QSO} & \colhead{\zabs} & \colhead{log~$N$(H~{\sc i})} & \colhead{W(Si~{\sc ii})} & \colhead{Z(P08)} & \colhead{Z} & \colhead{$\int \tau_{21} dv$} & \colhead{$T_s$} \\ 
\colhead{}     &  \colhead{}     &  \colhead{} & \colhead{(\AA)} & & \colhead{} & \colhead{(\kms)}    & \colhead{(K)} \\ 
\colhead{(1)} & \colhead{(2)} & \colhead{(3)} & \colhead{(4)} & \colhead{(5)} & \colhead{(6)} & \colhead{(7)} & \colhead{(8)} \\ 
}
\startdata
\vspace{-0.1cm} \\
M0416-3339 & 2.8561 & 20.50$\pm$0.10 &0.12$\pm$0.03 &$-$2.2 & $\le -2.3^a$& RFI & .... \\ 
M0610-2301 & 2.3987 & 20.85$\pm$0.10 &0.66$\pm$0.02 & $-$1.2 & $\le -1.1^b$&$\le 0.64$ & $\ge$603 \\ 
M1013-2546 & 2.6834 &20.35$\pm$0.15                &  $\le0.90$            &....        &  ....           &$\le 0.72$ &$\ge$169     \\ 
M1351-0129 & 2.7719 & 20.50$\pm$0.10 &0.19$\pm$0.03 &$-1.9$ &$\le -1.2^c$ & RFI & .... \\ 
M1619-0939 & 2.7934 & 20.55$\pm$0.10 & 1.03$\pm$0.05 & $-0.9$ & $\le -1.9^c$ & RFI & .... \\ 
M2154-3826 & 2.7613 & 21.00$\pm$0.10 &0.53$\pm$0.01&$-1.3$&$ -1.60\pm0.11^d$&.... & ....  \\ 
\label{tab:dla}
\enddata
\tablecomments{ 
Column 1: source name. Columns 2 and 3: DLA redshift and \hi\ column density. Column 4: Si~{\sc ii} equivalent width. Column 5: metallicity inferred using W(Si~{\sc ii}) and correlation from \cite{Prochaska08}. Column 6: metallicity measured using weak Si~{\sc ii} or S~{\sc i} lines.  Column 7: 5$\sigma$ 21-cm optical depth limit, considering $\Delta v$ = 25\,\kms. 
\\
$a$: Based on Si~{\sc ii}$\lambda1301$ line; $b$: Based on S~{\sc ii}$\lambda$1250 line; $c$: Based on Si~{\sc ii}$\lambda$1808 line; $d$: Using the curve of growth analysis.
}
\end{deluxetable*}

Our SALT spectra allow us to search for Fe~{\sc ii}$\lambda\lambda\lambda$2343, 2374, 2383 for \zabs$\le$2.15 and DLAs at $z>2.65$.
For this search, we complement our SALT-NOT survey spectra with more sensitive long-slit SALT observations of 25 quasars $\mathrm{z_{em} >2.7}$  obtained to search for extended \lya\ emission halos associated with powerful radio loud quasars \citep[see][for an example]{Shukla21}. 
In total, we identify 7 DLAs and 1 strong Fe~{\sc ii} absorber in our sample. 
Implications of lack of 21-cm absorption in these high \hi\ column density absorbers are discussed below. In the redshift range 2.15 $\le$ \zabs $\le$ 2.65 we also identify 21 C~{\sc iv} absorbers for which neither Fe~{\sc ii} nor DLA can be searched in our spectrum. The redshifted H~{\sc i} 21-cm line frequencies corresponding to these \civ\ absorbers are unaffected by RFI and no \hi\ absorption is detected. Note that \civ\ can trace a wide range of ionization stages and so is not a good indicator of the presence of a DLA or 21-cm absorption.  We will use this only as an indicator of the possible presence of multi-phase gas along the sight line.
%


First, we focus on the subset of 23 quasars that have sufficient SNR in optical continuum and, hence, suitable to search for \lya\ absorption.  The absorption profiles of 6 DLAs detected from this subsample are shown in Fig.~\ref{fig_dlafit}. The measured absorption redshifts are consistent with 5 of these being intervening systems and remaining one (\zabs = 2.7613 towards M2154-3826) being a PDLA.
%

We also detect strong \lya\ absorption at the systemic redshift of M0507$-$3624. While we see the evidence of the damping wings and a wide range of absorption from singly ionised species, we also see non-zero flux in the core of the \lya\ absorption line. It is possible that this system is similar to the associated H$_2$ bearing DLAs studied by \citet{Noterdaeme19} where the presence of flux in the absorption core is related to the partial coverage.
Based on damping wings we estimate $N$(H~{\sc i})$\sim 10^{20.2}$\,\cmsq. However, with the present spectra we are unable to confirm the origin of residual flux in the core. Higher spectral resolution dataset covering both \lya\ and Ly$\beta$ absorption is needed to get an accurate estimate of the $N$(H~{\sc i}) for this absorber. For the purpose of this paper, we consider this as a candidate PDLA.

For a total redshift path of 9.3, the detection of 5 intervening DLAs correspond to a number of DLAs per unit redshift, $n_{DLA}$ = 0.54 $\pm$ 0.24.  This is slightly higher but due to large uncertainties consistent with the measurement of 0.24$\pm$0.02 based on SDSS DLAs by \citet{Noterdaeme12dla}, and $0.26^{+0.06}_{-0.05}$ towards radio-loud quasars based on the combined CORALS and UCSD samples by \citet{Jorgenson06}. Since, the quasars in our sample are fainter than previous surveys (see Fig.~\ref{fig:samp}), it is also possible that there is indeed a dependence between $n_{\rm DLA}$ and the faintness of quasars as noted by \citet[][]{Ellison01} for the CORALS sample.

As discussed in next section, in comparison to associated \hi\ 21-cm absorption detection rates in low-$z$ AGNs, the detection of just 2 PDLAs and one associated \hi\ 21-cm absorber (M1540-1453) from our sample may seem surprisingly low.  But actually the detection of 3 PDLAs from our sample is in fact a factor of 3 larger than what we expect from the statistics of PDLAs observed at $z\sim 3$ in SDSS \citep[][]{Prochaska08}. Interestingly, from the statistics of damped H$_2$ absorption lines, \citet[][]{Noterdaeme19} have suggested that the PDLA fraction in \citet{Prochaska08pdla} may have been underestimated. 
A complete search of \lya\ absorption towards all the targets in our sample will confirm the above mentioned excesses of DLAs and PDLAs. Specifically, from the observations of all the sources ($\Delta z \sim$  60), we expect to detect another $\sim$30 DLAs.
%

Using SALT 1D spectra we measure redshift, $N$(H~{\sc i)} and rest equivalent widths of metal absorption lines corresponding to these DLAs. 
These measurements are provided in Table.~\ref{tab:dla}. The quoted error in $N$(H~{\sc i)} also include continuum placement uncertainties. The single component Voigt profile fits to the DLAs are shown in Fig~\ref{fig_dlafit}. The rest equivalent width of Si~{\sc ii}$\lambda$1526\AA\ lines are provided in column 4 of Table.~\ref{tab:dla}.
The metallicities inferred using the W(Si~{\sc ii}) and metallicity correlation \citep[see equation 1 of][]{Prochaska08} are given in column 5.
We also estimated metallicity using weak transitions of Si~{\sc ii} or S~{\sc ii}. These are provided in column 6.  
For \zabs = 2.7613 PDLA towards M2154$-$3826 we detect several weak transitions of Fe~{\sc ii}, Ni~{\sc ii} and Si~{\sc ii}. We used single component curve of growth to measure metallicities in this case. 
Overall, the metallicities of these DLAs are typically less than a tenth of the solar metallicity. However, given the poor spectral resolution of our data these estimates may suffer from hidden saturation effects.

Unfortunately strong persistent RFI at 360-380 MHz prevents \hi\ 21-cm line search at $z$ = 2.73-2.94 (see Fig.~\ref{fig:gz}). 
Thus, we could observe 21-cm line only for the \zabs = 2.3987 DLA towards M0610-2301 and \zabs = 2.6834 towards M1013-2546. We do not detect 21-cm absorption from these systems. The 5$\sigma$ integrated optical depths are provided in column 7 of Table~\ref{tab:dla}. 
M0610-2301 is a compact radio source in the modest quality VLASS quick look image.  The deconvolved source size is $1.3^{\prime\prime}\times 0.1^{\prime\prime}$ with a position angle of $180^\circ$ (i.e., size $<$ 11\,kpc at \zabs).  
M1013-2546 also appears to be a core-dominated source.  Thus, we assume complete coverage i.e., $f_c =1$.
This together with the observed $N$(H~{\sc i}) translates to a lower limit on the spin temperature, $T_S\ge 603$ K for M0610$-$2301 and $\ge$ 169K for M1013$-$2546. These limiting values of spin temperatures are higher than the measured median $N$(H~{\sc i}) weighted $T_S$ of 70\,K for the cold neutral medium (CNM) in our galaxy \citep[see][]{Heiles03}.
We note that \hi\ 21-cm observations of 23 DLAs from radio-selected samples of CORALS, UCSD and \citet[][]{Ellison08} are available in the literature \citep[][]{Srianand12dla, Kanekar14}.  The overall 21-cm absorption detection rate of 3/25 (12$^{+11}_{-7}$\%) is consistent with the measurements from optically selected samples and the conclusion that DLAs at $z>2$ are predominantly warm and tend to show high spin temperatures \citep[see also][]{Petitjean00}.
Much larger radio-selected surveys ($\Delta {\rm X} \gtrsim 10^4$) are needed to uncover the population of dusty DLAs.

Our SALT spectra also allow us to detect Fe~{\sc ii}$\lambda$2383 line for \zabs$\le 2.15$. We detect a strong Fe~{\sc ii} absorber at \zabs = 2.1404 towards M0652-3230. This system also shows absorption lines from other singly ionized species (i.e Si~{\sc ii} and  Al~{\sc ii}).  All these suggest high $N$(\hi) \citep[][]{Dutta17fe2}.  
But as can be seen from Fig.~\ref{fig:maps}, the radio emission is dominated by the double lobe structure. The separation between the two lobes is $\sim27^{\prime\prime}$ (250\,kpc at \zabs). 
Therefore, the radio and optical sight lines are well separated. This explains the non-detection of 21-cm absorption at the redshift of Fe~{\sc ii} absorbers. 

\section{Associated absorption statistics}    
\label{sec:assoc}  

\begin{figure} 
\centerline{\vbox{
\centerline{\hbox{ 
\includegraphics[trim = {0cm 0cm 0cm 0cm}, width=0.45\textwidth,angle=0]{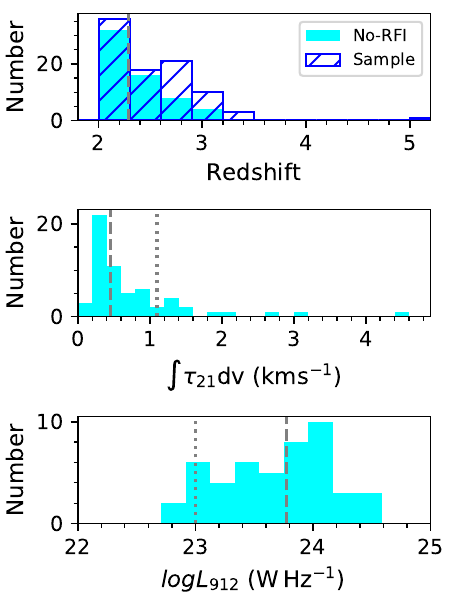}  
}} 
}}  
\vskip+0.0cm  
\caption{
Distributions of quasar redshift (hatched histogram for all targets and filled for those unaffected by RFI), 5$\sigma$ 21-cm optical depth limits and 912\AA\ luminosities for the AGNs searched for associated \hi\ absorption. 
The vertical dashed lines mark the median for each distribution.  
The dotted lines in middle and bottom panels correspond to $\int\tau_{21}$dv = 1.1\,\kms  and $L_{912} = 10^{23}$\,W\,Hz$^{-1}$.
} 
\label{fig:assoc}   
\end{figure} 

We searched for \hi\ 21-cm absorption within 3000\,\kms\ to the quasar emission line redshift. 
In 28/88 cases (32\%), the redshifted frequency is affected by RFI.  For remaining 60 sources, the distributions of redshift and 5$\sigma$ 21-cm optical depth limit for a width of 25\,\kms\ are shown in Fig.~\ref{fig:assoc}.  
The median redshift of this subsample which includes M1312-2026 ($z=5.064$) is 2.288. 
For sources with resolved morphology (Fig.~\ref{fig:maps}), we have searched for absorption towards multiple components but consider only the optical depth limit for the strongest component for statistics.

The only 21-cm absorption detection from the survey is described in Section~\ref{sec:dets}.  For non-detections, the optical depth limits span a range of 0.103 - 3.6\,\kms. 
The detection rate of $1.6^{+3.8}_{-1.4}$\% without any sensitivity cut is among the lowest for a 21-cm absorption line survey.
If we adopt the detection rate based on low-$z$ surveys, we would expect to detect approximately 20 \hi\ 21-cm absorbers from our sample.  For example, \citet[][]{Maccagni17} reported a 27\%$\pm$5\% detection rate from the observations of 248 AGNs at $0.02 < z < 0.25$. 
The 3$\sigma$ peak optical depth (resolution $\sim$ 16\,\kms) limits are typically between 0.005 - 0.2 (median $\sim$ 0.04).  For 25\,\kms\ width adopted by us, the range corresponds to 5$\sigma$ integrated optical depth, ${\cal{T}}$, of 0.2 - 7.1\,\kms (median $\sim$ 1.41). In our survey, we achieve  ${\cal{T}}$ better than 1.1\,\kms\ in 90\% of the cases.   
Thus, stark contrast in detection rates is not due to differences in the sensitivity.  
%

At low-$z$, the vast majority of targets have WISE color $W1 - W2$, which is an indicator of AGN accretion rate, to be less than 0.6. Among these the higher fraction of detections are associated with larger $W2 - W3$ colors, which is an indicator of star formation rate \citep[][]{Glowacki17, Chandola20}.  The compact radio sources still embedded within such gas and dust rich environments are well placed to exhibit 21-cm detections.  
The powerful quasars identified by our MIR-wedge (see equation~\ref{eqwedge}) do not overlap with the above-mentioned low-$z$ AGNs in color space defined by $W1 - W2$ and  $W2 - W3$.
Thus, we hypothesize that the low non-detection rate in our sample is a consequence of the gas poor environment of host galaxies.  Since the detectability of \hi\ 21-cm absorption also depends on the thermal state of the gas and the morphology of radio emission, we validate this hypothesis through \lya\ and metal lines covered in our SALT-NOT spectra.

First, we consider following four sources\footnote{We exclude M1312-2026 at $z=5.064$ as all the lines except \lya\ are redshifted into IR and not covered in our SALT spectrum.} at 2.7$< z <$3.5 : M0507$-$3624, M1013$-$2546, M1351$-$1019 and M2047$-$1841. For these we have access to both \lya\ and metal absorption lines and the 21-cm frequency is not affected by RFI.
In the cases of M1013$-$2546 and M2047$-$1841, we do not detect a PDLA or any associated \civ\ absorption system. This suggests the absence of neutral and metal enriched ionized gas along the sight lines.
In the case of  M0507$-$3624, we identified a potential PDLA. For $N$(H~{\sc i})$\sim 10^{20.2}$\,\cmsq\  inferred from the damping wing of \lya\ absorption and \hi\ 21-cm non-detection,  we place a lower limit of 216\,K for the spin temperature ($f_c = 1$). 
In the case of M1351$-$1019 while \lya\ absorption is not prominent we do detect associated \civ\ absorption. In this case, we also detect extended \lya\ emission. All this suggests that these two sources are associated with gas rich environment.  Therefore, the lack of \hi\ 21-cm absorption here may just be due to lower CNM filling factor. 

In general, 22 i.e., $\sim$23\% quasars in our sample show associated \civ\ absorption within 3000 \kms\ to the \zem.
In 11 of these cases 21-cm absorption could not be searched due to strong RFI.
Among remaining 10, we also detect  H~{\sc i} \lya\ absorption in 4 cases but none are DLAs. 
In the remaining 6 cases, the \lya\ absorption is not covered in the spectra but we do not detect corresponding absorption due to any low-ionization species such as Fe~{\sc ii} (in 4 cases) and Si~{\sc ii} (in 6 cases). 
Since \civ\ may come from a wide range of ionization stages, the absence of strong \lya\ and low-ionization absorption lines indicate the lack of sufficient high column density neutral gas along the line of sight. 

From the above we conclude that the high fraction of quasars in our sample are indeed residing in gas and dust poor environments. 
An interesting counterpoint to the lack of \hi\ 21-cm absorption in our sample is provided by the 100\% detection rate of molecular gas through CO emission in a sample of eight hyper-luminous WISE/SDSS (WISSH) quasars at $z\sim2-4$ \citep[][]{Bischetti21}. We note that a majority (6/8) of these would be selected by our MIR-wedge (Equation~\ref{eqwedge}). But WISSH quasars are much brighter (1.5 Vega magnitude compared to our sample) in the $W4$ band (22$\mu$m) of WISE.  The deliberate selection of WISSH quasars as most luminous infrared sources ensures that they are being observed through dust clouds \citep[][]{Weedman12}, and perhaps represent an early phase in the evolution of quasar.  Approximately only $\sim$10\% of quasars in our sample have $W4$ magnitudes comparable to above-mentioned WISSH quasars with CO detections, and only in 3 cases ${\cal{T}} <$ 1.1\,\kms\ i.e., the sensitivity to detect CNM in $N$(\hi)$>10^{20}$\,\cmsq\ is achieved. 
Although this may seem to conflict with our MIR-selected sample strategy but luminous quasars only spend a small fraction of their total lifetime \citep[$\sim10^7$\,yr;][]{Martini01} in the dust-obscured phase.  Therefore, the the representation of dust-obscured quasars in our unbiased sample will also be proportionately small.  
Considering only the AGNs with sensitivity to detect CNM in $N$(\hi)$>10^{20}$\,\cmsq\ gas, we estimate the CNM covering factor in the unobscured phase of quasars with radio luminosity $L_{\rm 1.4\,GHz} \simeq 10^{27 - 29.3}$\,W\,Hz$^{-1}$ to be 0.02. 

Although, the most straightforward explanation for low detection rate in our sample is gas and dust poor environment at host galaxy scales but the detectability of gas towards the AGN may also be influenced by the additional factors such as high intrinsic radio or ultraviolet luminosities of AGN \citep[][]{Curran10assoc, Aditya16, Grasha19} and the distribution of gas at nuclear scales \citep[][]{Gupta06uni}.  Since high $L_{UV}$ merely helps to select core-dominated AGNs, these two are in fact interlinked in the context of AGN unification scheme.  We estimate the 912\AA\ luminosities of the AGNs searched for associated \hi\ 21-cm absorption by interpolating the photometry from PS1.  The distribution of $L_{912}$ is shown in the bottom panel of Fig.~\ref{fig:assoc}. Clearly, the majority of objects have $L_{912} > 10^{23}$\,W\,Hz$^{-1}$, where the associated \hi\ 21-cm absorption is rarely detected.
%

Unfortunately, none of the CO-detected quasars from \citet[][]{Bischetti21} are bright enough at radio wavelengths to search for \hi\ 21-cm absorption.  
Interestingly, their available SDSS spectra do not show the presence of high-column density neutral gas i.e., a PDLA along the quasar sight line (although ionized gas outflows are present). 
Thus, although the molecular gas is distributed in rotating disks  \citep[extent 1.7 - 10\,kpc;][]{Bischetti21}, it is oriented such that cold gas cross-section towards the quasar sight line is minimal. A spatially extended few kpc-sized radio source embedded in such an environment may have still shown \hi\ 21-cm absorption (e.g., refer to the cases of B2\,0902+345 and MG\,J0414+0534 in Section~\ref{sec:dets}).

Finally, we note the non-detection towards the highest redshift quasar (\zem = 5.062) in our sample: M1312-2026 \cite[see also][for 21-cm non-detections towards two $z\sim$5 AGNs]{Carilli07}.  This brightest radio loud quasar at $z>5$, has a  radio-loudness parameter of 
R = $f_{\nu,5GHz}$/$f_{\nu,4400\AA}$ = $1.4\times10^4$. This R value is an order-of-magnitude greater than that of any other $z > 5$ AGN known-to-date \citep[][]{Momjian18, Saxena18}.  The host galaxies of quasars at such high redshifts can be associated with large amounts of dust and molecular gas ($>10^{10}\,{\rm M_\odot}$), and high inferred star formation rates ($>100\,{\rm M_\odot\,yr^{-1}}$) \citep[][]{Venemans17, Decarli18, Feruglio18}.
Our \hi\ 21-cm non-detection corresponds to a 5$\sigma$ upper limit of $N$(\hi)$< 4\times 10^{19}$\,\cmsq  = 100\,K; $f_c$ = 1 assumed) but can miss narrow absorption components due to the RFI.  The current SALT spectrum also covers only \lya\ emission.  
Further investigation on the nature of this very intriguing non-detection require IR spectra and sub-arcsecond scale radio imaging which are in progress.

\section{Summary and outlook}    
\label{sec:summ}  

This paper described a spectroscopically blind search for \hi\ 21-cm absorption lines in the wide band uGMRT spectra of 88 AGNs at $2 < z < 5.1$.
We also applied the same formalism to constrain the occurrence of intervening OH 18-cm main lines.
We show that compared to previous radio-selected samples of quasars to search for DLAs, our sample for the uGMRT survey has targeted fainter objects (median $i$ = 19.5\,mag; see Fig.~\ref{fig:comp}) and is a close representation of the underlying luminosity function of quasars.  Thus, our dust-unbiased sample of AGNs with median radio spectral index, $\alpha^{1.4}_{0.4}$ = $-0.38$, redshift, $z$ = 2.5 and spectral luminosity, $L_{\rm 1.4GHz}$ = $10^{27-29.3}$\,W\,Hz$^{-1}$ is ideally suited to determine the occurrence of cold atomic gas ($T\sim$100\,K) towards powerful quasars at $z>2$.  

Through a spectroscopically blind search of absorption lines in all the uGMRT spectra, we detected one new associated \hi\ absorption which is towards M1540-1453 at \zabs = 2.1139.  No intervening \hi\ 21-cm absorption line is detected. 
Our detection is only the fourth associated \hi\ 21-cm absorber known at $z>2$.  It has a \hi\ column density of $(2.06\pm0.01)\times 10^{21} ({T_S\over 100}) ({1\over f_c})$ cm$^{-2}$.  In our SALT spectrum, the peak of \civ\ emission and low-ionization metal absorption lines are coincident with that of the 21-cm absorption.  The overall properties of 21-cm absorption are consistent with it originating from a circumnuclear disk or gas clouds associated with the host galaxy. 
The CO emission line observations and optical spectra covering the \lya\ absorption (i.e., $\lambda\sim3785$\AA) along with the sub-arcsecond scale imaging will allow us to understand the origin of cold gas detected in 21-cm absorption.

Our survey is sensitive to detect CNM in DLAs corresponding to a total redshift and comoving path length of $\Delta z = $ 38.3 and  $\Delta X =$ 130.1, respectively.  Using this we constrain the incidence or number of 21-cm absorbers per unit redshift and comoving path length to be $n_{21} <$  0.048 and $\ell_{21} < $  0.014, respectively. 
The same formalism applied to OH main line at 1667.359\,MHz corresponds to total redshift and comoving path length of $\Delta z = $ 44.9 and  $\Delta X = $ 167.7, respectively. The number of OH absorbers per unit redshift and comoving path length are $n_{\rm OH} < $ 0.041 and $\ell_{\rm OH} < $  0.011, respectively. We note that the number of DLAs per unit redshift interval i.e., $n_{\rm DLA}$($z$) at $2.3\le z \le 2.9$ is in the range of 0.21 to 0.29 \citep[][]{Noterdaeme12dla}. 
This implies that the covering factor of CNM gas in DLAs is $\le$20 percent.
These upper limits are also consistent with $n_{\rm CNM}$ = 0.012 estimated using H$_2$ and C~{\sc i} absorbers, also tracers of cold gas, at high-$z$ \citep[][]{Krogager20}.  
Our result shows that a moderately larger survey such as MALS with $\Delta$X$ \gtrsim 10^{4}$ is important to precisely characterise the CNM fraction and its redshift evolution at high-$z$.

The low-$z$ AGNs ($z<0.25$) exhibit \hi\ 21-cm absorption detection rates of $\sim$30\% \citep[e.g.,][]{Maccagni17}.  Compared to this the low associated \hi\ 21-cm absorption detection rate ($1.6^{+3.8}_{-1.4}$\%) and the CNM filling factor of 0.2 from our survey is intriguing. We show that this is most likely due to the fact that the powerful quasars in our sample are residing in gas and dust poor environments, and that luminous quasars only spend a small fraction of their total lifetime in dust-obscured phase.  
We use the spectral coverage of \lya\ and various metal absorption lines in our optical spectra to confirm the absence of high column density atomic gas towards the quasar sight lines.  

From our SALT spectra, we report detections of 5 intervening DLAs and 2 PDLAs in our sample.  The measured number of DLAs per unit redshift, $n_{\rm DLA}$ = 0.54 $\pm$ 0.24 is slightly higher but due to large uncertainties consistent with the measurement based on SDSS DLAs \citep[][]{Noterdaeme12dla} and the combined CORALS and UCSD sample of radio-selected quasars \citep[][]{Jorgenson06}.  Interestingly, the PDLA detection fraction is also a factor of 3 larger.  Since the quasars in our sample are fainter than in the previous surveys, there may indeed be a dependence between $n_{\rm DLA}$ and optical faintness of quasars \citep[][]{Ellison01}. 
These results also underline the need for larger surveys of dust-unbiased DLAs. 
Due to limited spectral coverage, we could search for \lya\ in only 30\% of our SALT-NOT sample presented here.  A complete search of \lya\ absorption towards all the targets will allow us to examine the above-mentioned excesses at a higher significance level.  

Eventually, much larger radio-selected surveys ($\Delta$X $\gtrsim10^4$) such as MALS are needed to uncover the population of dusty DLAs. The key science objectives are summarized in \citet[][]{Gupta17mals}, and the survey is well underway.  The first L- and UHF-band spectra based on the science verification data are presented in \citet[][]{Gupta21} and \citet[][]{Combes21}, respectively. 
Each MALS pointing is centered at a bright ($>200$\,mJy at 1\,GHz) radio source.  Through wideband spectra of the central bright radio source and the numerous off-axis radio sources, it will sample the column density distribution ($N$(\hi)$>5\times10^{19}$\,\cmsq; $T_s$ = 100\,K) relevant to characterize the cross-section of cold atomic gas in and around normal galaxies and AGNs at $0<z<1.4$.  
Simultaneously, it will also be sensitive to detect OH main line absorption at $z<1.9$ in gas with $N$(OH)$>2.4\times10^{14}$\,\cmsq\ (excitation temperature = 3.5\,K).
Since the formation of OH is tightly coupled to H$_2$, the measurement of OH cross-section at $z<2$ will be a crucial input through which to understand the redshift evolution of CNM cross-section \citep[][]{Balashev21}. 

The work presented here is the first to examine the detectability of high-$N$(\hi) absorbers in a MIR selected sample of quasars with searches based on radio / optically selected quasars.  Although only one associated \hi\ 21-cm absorber is detected, the slight excess of DLAs and PDLAs in our sample is encouraging and with larger surveys could potentially lead to the estimates of dusty AGNs missed in optically selected surveys \citep[the fraction could be anywhere between 10-50\%;][]{Richards03, Glikman12}.  Some of the ongoing work we are pursuing is to understand the impact of MIR selection on the colors of selected quasars.  In the near future, we will also present the complete census of DLAs and PDLAs in our sample, and results from the targeted deeper \hi\ 21-cm line observations towards these. As such, this paper presents a crucial step towards a detailed understanding of the high-$N$(\hi) searches at $z>2$ to be eventually taken up with the low-frequency component of the SKA \citep[][]{Morganti15}.

\acknowledgments

We thank the anonymous referee for helpful suggestions.
We thank GMRT staff for their support during the observations.  
GMRT is run by the National Centre for Radio Astrophysics of the Tata Institute of Fundamental Research.
This  work  is  based  on  observations  made  with  SALT and NOT.  
The uGMRT data were processed using the MALS data processing facility at IUCAA.
The CASA package is developed by an international consortium of scientists based at the National Radio 
Astronomical Observatory (NRAO), the European Southern Observatory (ESO), the National 
Astronomical Observatory of Japan (NAOJ), the Academia Sinica Institute of Astronomy 
and Astrophysics (ASIAA), the CSIRO division for Astronomy and Space Science (CASS), 
and the Netherlands Institute for Radio Astronomy (ASTRON) under the guidance of NRAO. 
The National Radio Astronomy Observatory is a facility of the National Science Foundation operated under cooperative agreement by Associated Universities, Inc.

\facilities{NOT, SALT and uGMRT}

\software{ARTIP \citep[][]{Gupta21}, Astropy \citep[][]{Astropy13, Astropy18}, CASA \citep[][]{Mcmullin07} and Matplotlib \citep[][]{Hunter07}. }

\def\aj{AJ}%
\def\actaa{Acta Astron.}%
\def\araa{ARA\&A}%
\def\apj{ApJ}%
\def\apjl{ApJ}%
\def\apjs{ApJS}%
\def\ao{Appl.~Opt.}%
\def\apss{Ap\&SS}%
\def\aap{A\&A}%
\def\aapr{A\&A~Rev.}%
\def\aaps{A\&AS}%
\def\azh{AZh}%
\def\baas{BAAS}%
\def\bac{Bull. astr. Inst. Czechosl.}%
\def\caa{Chinese Astron. Astrophys.}%
\def\cjaa{Chinese J. Astron. Astrophys.}%
\def\icarus{Icarus}%
\def\jcap{J. Cosmology Astropart. Phys.}%
\def\jrasc{JRASC}%
\def\mnras{MNRAS}%
\def\memras{MmRAS}%
\def\na{New A}%
\def\nar{New A Rev.}%
\def\pasa{PASA}%
\def\pra{Phys.~Rev.~A}%
\def\prb{Phys.~Rev.~B}%
\def\prc{Phys.~Rev.~C}%
\def\prd{Phys.~Rev.~D}%
\def\pre{Phys.~Rev.~E}%
\def\prl{Phys.~Rev.~Lett.}%
\def\pasp{PASP}%
\def\pasj{PASJ}%
\def\qjras{QJRAS}%
\def\rmxaa{Rev. Mexicana Astron. Astrofis.}%
\def\skytel{S\&T}%
\def\solphys{Sol.~Phys.}%
\def\sovast{Soviet~Ast.}%
\def\ssr{Space~Sci.~Rev.}%
\def\zap{ZAp}%
\def\nat{Nature}%
\def\iaucirc{IAU~Circ.}%
\def\aplett{Astrophys.~Lett.}%
\def\apspr{Astrophys.~Space~Phys.~Res.}%
\def\bain{Bull.~Astron.~Inst.~Netherlands}%
\def\fcp{Fund.~Cosmic~Phys.}%
\def\gca{Geochim.~Cosmochim.~Acta}%
\def\grl{Geophys.~Res.~Lett.}%
\def\jcp{J.~Chem.~Phys.}%
\def\jgr{J.~Geophys.~Res.}%
\def\jqsrt{J.~Quant.~Spec.~Radiat.~Transf.}%
\def\memsai{Mem.~Soc.~Astron.~Italiana}%
\def\nphysa{Nucl.~Phys.~A}%
\def\physrep{Phys.~Rep.}%
\def\physscr{Phys.~Scr}%
\def\planss{Planet.~Space~Sci.}%
\def\procspie{Proc.~SPIE}%
\let\astap=\aap
\let\apjlett=\apjl
\let\apjsupp=\apjs
\let\applopt=\ao
\bibliographystyle{aasjournal}
\bibliography{mybib}

\end{document}